\documentclass[twocolumn,
         showpacs,
         amsmath,
         amssymb,
         amsfonts,
         superscriptaddress,
         aps,
         prb,
         longbibliography,
         floatfix,
]{revtex4-2}
\usepackage[english]{babel}
\usepackage[utf8]{inputenc} 
\usepackage{dsfont}
\usepackage{SIunits}
\usepackage{hyperref}% add hypertext capabilities
\usepackage{graphicx}     % Allows to insert images from outside
\usepackage{color}

\begin{document}
\title{Photo-assisted current in the fractional quantum Hall effect as a probe of the quasiparticle operator scaling dimension}
\author{Bruno Bertin-Johannet}
\email{bruno.bertin@cpt.univ-mrs.fr}
\author{Laurent Raymond}
\author{J\'er\^ome Rech}
\author{Thibaut Jonckheere}
\author{Beno\^it Gr\'emaud}
\affiliation{Aix Marseille Univ, Universit\'e de Toulon, CNRS, CPT, IPhU, AMUtech, Marseille, France}
\author{D. Christian Glattli}
\affiliation{Universit\'e Paris-Saclay, CEA-CNRS, SPEC, CEA-Saclay, Gif-Sur-Yvette, France}
\author{Thierry Martin}
\affiliation{Aix Marseille Univ, Universit\'e de Toulon, CNRS, CPT, IPhU, AMUtech, Marseille, France}

\begin{abstract}
We study photo-assisted transport for the edge states of a two dimensional electron gas in the fractional quantum Hall regime, pinched by a single quantum point contact. We provide a general expression of the photo-assisted current using a Keldysh-Floquet approach,
when the AC drive is applied either directly to the edge states, or when it modulates the tunneling amplitude
at the quantum point contact. Strikingly, for a simple cosine modulation of the tunneling amplitude, the phase shift of the second harmonic of the photoassisted current is directly related to the scaling dimension of the quasiparticle operators describing the fractional excitations.
As the scaling dimension is intimately related to the statistics, our proposal of a gate modulation of the backscattered current provides a diagnosis of the statistics of Laughlin quasiparticles using a simple quantum point contact geometry.

\end{abstract}
\maketitle
\section{Introduction}\label{sec:intro}
The Fractional Quantum Hall Effect (FQHE) is a strongly correlated state of a two dimensional electron gas (2DEG) under strong magnetic field in the presence of Coulomb interaction. It has generated considerable interest on the theoretical and the experimental side since its discovery \cite{tsui82,laughlin83}. A key feature resides in the fact that excitations in the FQHE are quasiparticles which carry a fractional charge and which bear fractional statistics, which is intermediate between that of fermions and of bosons, thus the terminology ``anyons''. In the pioneering work of Ref.~\cite{laughlin83} this new state of matter was shown to appear for filling factors $\nu$ equal to the inverse of an odd number. The fractional charge is then specified by $e^*=\nu e$ and the statistical angle equals $\pi\nu$.

It is therefore challenging to identify an experiment which measures the statistical angle of anyons independently of their charge. The charge of anyons has been successfully identified via the measurement of the Fano factor \cite{kane94,chamon95,saminadayar97,depicciotto97}, or via the
identification of the Josephson frequency $\omega_J=e^*V/\hbar$ which is accessed either through a photo assisted transport noise measurement at zero frequency \cite{crepieux04,kapfer19}, or through a direct finite frequency noise measurement \cite{bisognin19}. All the above experiments were performed in the weak backscattering regime of a single quantum point contact (QPC), where the accepted paradigm is that quasiparticle exchange between two opposite edge states constitute the dominant tunneling process. 

The detection of fractional statistics has proved to represent an even greater challenge. On the theoretical side, several proposals considered setups with several QPCs in a Hanbury Brown and Twiss geometry \cite{hanburybrown56} where noise crossed correlations were computed \cite{safi01,vishveshwara03,kim05,kim06}. Alternatively, more recently, proposals using either Fabry Perot interferometry or Hong-Ou-Mandel collisions of quasiparticles were put forward \cite{halperin11,rosenow16,feldman21}. 
These proposals generated considerable attention from the experimental community as illustrated by two recent pioneering experiments with a strong claim that the measurement of the statistical angle has been achieved \cite{bartolomei20,nakamura20}.
Nevertheless, the detection of the statistical angle of anyons continues to generate a lot of excitement as illustrated by recent theoretical works involving anyon braiding \cite{lee19,lee20,morel22,jonckheere22,mora22,schiller23}.

One obvious drawback of both theoretical proposals and experimental detection schemes for the statistical angle of quasiparticles resides in the fact that the setups typically involve several QPCs, which constitute both a theoretical and an experimental challenge, and typically a (rather involved) noise measurement. Here we address the issue whether some signatures of the statistics would already be present in a measurement of some specific features of the time dependent current.

%%%%%

Like in any edge theory, the quantum numbers characterizing the local quasiparticle excitations are an essential ingredient. The focus is typically drawn on two of these: the effective charge and the scaling dimension (i.e. the exponent which characterizes the time decay of the quasiparticle operator correlation function). In practice, the statistical angle of anyons is intimately tied to the scaling dimension of the quasiparticle operator. Indeed, in the hydrodynamical picture of the FQHE of Ref.~\onlinecite{wen92}, which characterizes the edge excitations of a FQHE fluid by a chiral Luttinger liquid model, the scaling dimension turns out to provide an upper bound for the statistical angle, the two being equal when all the modes of a given edge have the same chirality, up to a factor $\pi$. The present work relies on the general framework of the chiral Luttinger liquid (CLL) model~\cite{wen92}, which provides us with a low-energy effective theory describing the physics of the edge states and their excitations. While unable to offer a perfectly realistic detailed description, as one would expect from a low-energy theory,  it works exceptionally well in most cases, capturing universal features that can be observed in actual experiments. 

Historically, the CLL model was initially used in Refs.~\onlinecite{kane94, chamon95} to predict that the transport properties of a biased QPC in the FQHE could be used to detect the fractional charge of quasiparticle excitations. This prompted the seminal experimental works of Refs.~\onlinecite{saminadayar97} and \onlinecite{depicciotto97} to actually measure the Fano factor in such a system and confirm theoretical predictions, thus providing us with the first experimental proof of the anyonic fractional charge. These measurements of the fractional charge were later confirmed by the the detection of the Josephson frequency in photo-assisted shot noise~\cite{kapfer19} and the measurement of high-frequency noise generated by the tunneling of quasiparticles~\cite{bisognin19}, both in line with the corresponding predictions from the CLL theory~\cite{chamon95, crepieux04, bena07}. Since these first successes from the mid-nineties, the CLL has become the default theoretical description of edge state physics and every experimental result on quantum transport in the FQHE has been confronted with the CLL model prediction, most of the time with success.  This is also true of the latest, most impactful experimental works carried out in the field, which claim to probe the anyonic statistics of quasiparticle excitations~\cite{bartolomei20,nakamura20,lee23} and whose findings are all reproduced by theoretical approaches based on the CLL model.

Naively speaking, one could argue that the scaling dimension of anyons should be accessible via a DC transport measurement in the weak backscattering regime of a single QPC, provided that one measures carefully the backscattering DC current-voltage characteristics. This should indeed manifest as a power-law (up to potential finite temperature effects~\cite{martin05} which may complicate the picture), whose exponent is directly related to the scaling dimension. Unfortunately, after various experiments have attempted to study the tunneling between DC biased edge states, the results are inconclusive: a power-law is often observed at low temperature but the exponent can vary substantially. It is now generally accepted that  the dependence of the tunneling current on the applied bias voltage in electrostatically confined QPCs  can significantly deviate from the theoretically predicted behavior of the CLL model (see e.g. \cite{grayson98, roche02, roddaro03, chang03, chung03, roddaro05,heiblum06} and references therein).  The simplest explanation usually put forward to explain this discrepancy has to do with electrostatic effects: changing the bias voltage between edge states modifies the electrostatics of the QPC, affecting its shape and the corresponding tunneling amplitude. This ultimately results in an alteration of the tunneling current as a function of the applied bias voltage, in an unknown nonuniversal way.

This, however should not cast doubts on the reliability of the CLL model to describe actual experimental platforms, as there are ways to limit these nonuniversal effects. One possibility to circumvent these issues is to analyze simultaneously the tunneling current and its noise, as the ratio of the two should be rid of most of the nonuniversal effects (those resulting in an unknown dependence of the tunneling amplitude with the bias voltage). This is precisely the method used in Refs.~\onlinecite{saminadayar97,depicciotto97} to achieve a reliable determination of the quasiparticle effective charge a few decades ago. This realization also lead to various proposals for extracting the scaling dimension via a crossed analysis involving the tunneling current along with the zero-frequency noise~\cite{snizhko15}, finite-frequency noise~\cite{ferraro14b} or thermal tunneling noise \cite{schiller22}, for more complicated states (featuring counterpropagating modes or non-Abelian statistics).

Another direction is proposed here, which relies on  a measurement of the sole current, rather than in conjunction with the noise, which is by itself more challenging to measure. In this paper, we argue that the scaling dimension of the quasiparticle operator could in principle be detected via the careful measurement of a phase shift of the photo-assisted current. Photo-assisted transport (PAT), i.e., electric current in the presence of an additional time dependent drive, typically achieved by shining microwaves on an otherwise DC voltage biased device, was pioneered in Ref.~\cite{tien63}. It has provided condensed matter physicists with new tools to probe fundamental properties of physical systems. Early proposals in mesoscopic devices considered normal metal systems \cite{lesovik94,schoelkopf98}, hybrid superconducting systems \cite{lesovik99,torres01,kozhevnikov00}, and the FQHE \cite{crepieux04,bisognin19}.
More recently photo-assisted transport has gained importance in the field of electron quantum optics, where a minimal excitation states, dubbed ``Levitons'' \cite{levitov96,dubois13a,dubois13b,kapfer19,bertin-johannet22} and containing integer electron charge, can be designed to generate a pure electron excitation above the Fermi sea, devoid of unwanted electron hole pairs.

%%%%%

In the present context of a single, DC voltage biased QPC in the weak backscattering regime of the FQHE, PAT can be envisioned in two different ways. Either one varies the gate voltage applied to the QPC, thus modulating the tunnel amplitude for quasiparticles between the two edge states, or an AC signal is added to the voltage drive applied to the two edge states. Interestingly, in the FQHE, to our knowledge, attention has mostly focused on the latter scenario. Only a few works have considered the effect of an  AC gate drive in the context of quantum Hall~\cite{safi10}, quantum splin Hall~\cite{ferraro13}, or in generic interacting mesoscopic devices~\cite{safi19}.

The first important result of the present work is to show that both types of drives (gate drive or voltage drive), despite their fundamental differences, can be described analytically with the same unified formalism. Unfortunately, for both types of drives, the measurement of the PAT current averaged over a period of the drive, does not provide any striking dependence on the scaling dimension of quasiparticles. Nevertheless, we point out that a complete description of the PAT current requires the knowledge of all of its harmonics at the AC drive frequency. In practice, this is precisely where the two types of drive lead to significantly different results.

Quite importantly, it turns out that the gate drive is the most suited of the two to detect the scaling dimension of FQHE quasiparticles.
The central result of this work is indeed to show that for a simple sinusoidal modulation of the gate this scaling dimension is directly accessible via the measurement of the second harmonic of the PAT current (conversely, for an AC voltage drive, we find that a similar analysis seems difficult to achieve). 
We thus find an appreciable range of parameters where the scaling dimension -- which is tied to the statistical angle -- can be easily extracted from a phase shift of the second harmonic of the current.

The paper is organized as follows. 
In Sec.~\ref{sec:model} we introduce the model for the FQHE and recall its main properties. 
The current operator through the QPC is derived and we show that the computation of the current under a periodic voltage drive and a periodic gate drive can be carried out with the same formalism.
In Sec.~\ref{sec:currentcomp} we find a general analytic formula of the backscattered current as a function of time that holds for arbitrary values of $\nu>0$.
Sec.~\ref{sec:weakbs} is devoted to finding the formula of the backscattered current in the weak backscattering regime.
The harmonics structure of this current in the weak backscattering regime is studied in Sec.~\ref{sec:harmonics} depending on the type of drive. 
In particular, we show the specificity of the harmonic gate drive which allows to isolate the filling factor in the second harmonic of the current.
In Sec.~\ref{sec:fermiliquid}, we show a computation of the backscattered current in the Fermi liquid limit through two different routes, which allows to check the consistency of our treatment.
Then, we compute the current in the strong backscattering regime in Sec.~\ref{sec:strongbs}, showing analytically its particular behavior.
Finally we conclude in Sec.~\ref{sec:conclusion} and propose further research tracks.

\section{Luttinger liquid basics}\label{sec:model}

\subsection{Model}

We propose a brief summary of the hydrodynamical approach to the description of the FQHE put forward by Wen in Ref.~\cite{wen92}.
In this approach, edge excitations are identified as surface waves of an incompressible irrotational two dimension quantum liquid with a perpendicular magnetic field $B$.
We restrict ourselves to Laughlin states of FQHE, see Ref.~\cite{laughlin83}, which amounts to assuming that there is only one fractional edge state. 

Defining the filling factor $\nu$ as the fraction of the lowest Landau level which is filled, the Hamiltonian for the FQHE describing the edge excitations reads:
\begin{equation}
	\mathcal{H}=\frac{v}{4\pi}\int_0^{L}\mathrm{d}x\, \left[\partial_x\phi(x)\right]^2\, ,
\end{equation}
where $v$ is the drift/Fermi velocity along the edges and $\phi(x)$ is a bosonic field which is related to the one dimensional electron density $\rho(x)=\sqrt{\nu}\partial_x\phi(x)/2\pi$.
Using the fermion anticommutation relation we establish a relation between the electron creation operator $\Psi^\dagger$ and the latter's density $\rho(x)$:
\begin{equation}
	\left[\rho(x),\,\Psi^\dagger(x')\right]=\delta(x-x')\Psi^\dagger(x)\, .	
\end{equation}
This means that the electron creation operator can be written as
\begin{equation}\label{eq:elop}
	\Psi^\dagger(x)=\frac{1}{\sqrt{2\pi a}} e^{\frac{i}{\sqrt{\nu}}\phi(x)}\, ,	
\end{equation}
where $a$ is a short distance cutoff.
Using the Kac-Moody commutation relation 
\begin{equation}
[\phi(x),\phi(x')]= -i\pi \text{sgn}(x-x')
\end{equation}
and the Baker Campbell Hausdorf identity, it can be shown that:
\begin{equation}
	\Psi(x)\Psi(x')=e^{-i\frac{\pi}{\nu}\text{sgn}(x-x')}\Psi(x')\Psi(x)\, ,
\end{equation}
so electron operators obey anticommutation relations only if $\nu=1, 1/3, 1/5, 1/7,...$, i.e., for Laughlin filling factors.
Note that other rational values of $\nu$ are physically attainable and exhibit the FQHE, but the above derivation needs to be generalized in order to include the presence of several bosonic fields \cite{wen92}. In what follows, we choose to focus on the simpler situation of the Laughlin series for clarity sake. The main results can be readily obtained for the case of a general Abelian FQH edge involving multiple bosonic modes, but do not fundamentally differ from the Laughlin filling factors.
In Appendix \ref{app:generalFQH}, we provide some elements of the model and main derivations for this more general case.

%Using bosonization techniques, see Ref.~\cite{vondelft98}, the quasiparticle operator has to be a vertex operator, furthermore it should respect the mutual exclusion principle with electrons, it therefore reads
The quasiparticle operator is a local vertex operator and it is required to commute with the electron operator, which justifies the choice:
\begin{equation}\label{eq:quasiparticle_operator}
	\psi(x)=\frac{1}{\sqrt{2\pi a}} e^{i\sqrt{\nu}\phi(x)}\, .
\end{equation}
where $a$ is a short distance cutoff. At zero temperature, the resulting correlation function, taken at position $x=0$, follows a power-law decay in time as
\begin{equation}
    \left\langle  \psi(0,\tau)\psi^\dagger (0,0)\right\rangle
    = e^{\nu \mathcal{G}(\tau)}\sim \tau^{-\nu}\, ,
\end{equation}
where, as explained in Appendix \ref{app:keldyshgf}, $\mathcal{G}(\tau)$ is the chiral bosonic field Green's function Eq.~\eqref{chiral_green}. This allows us to define the scaling dimension $\nu_D$ of the quasiparticle operator~\footnote{There exists an alternate definition in the literature, which only differs by an extra factor $1/2$. It is mostly a matter of convention and does not affect our main message and results.}, which, in the case of Laughlin filling factors, reduces to $\nu_D = \nu$.

From Eq.~\eqref{eq:quasiparticle_operator}, it is also possible to define the statistical angle $\Theta$. Indeed, focusing on a given time $\tau=0$, one readily obtains a nontrivial phase factor when exchanging two quasiparticles in real space, namely
\begin{equation}
	\psi(x)\psi(x')=e^{i\pi\nu\text{sgn}(x-x')}\psi(x')\psi(x)\, ,
\end{equation}
which is a clear illustration of anyonic statistics, with a statistical angle specified by $ \Theta = \pi \nu_D = \pi \nu$.

Note that so far, the fractional charge $e^*=\nu e$ has not been discussed, at it typically appears in discussions where electromagnetic fields/voltage biases are involved. 

\subsection{QPC current operator}

\begin{figure}
	\includegraphics[width=0.45\textwidth]{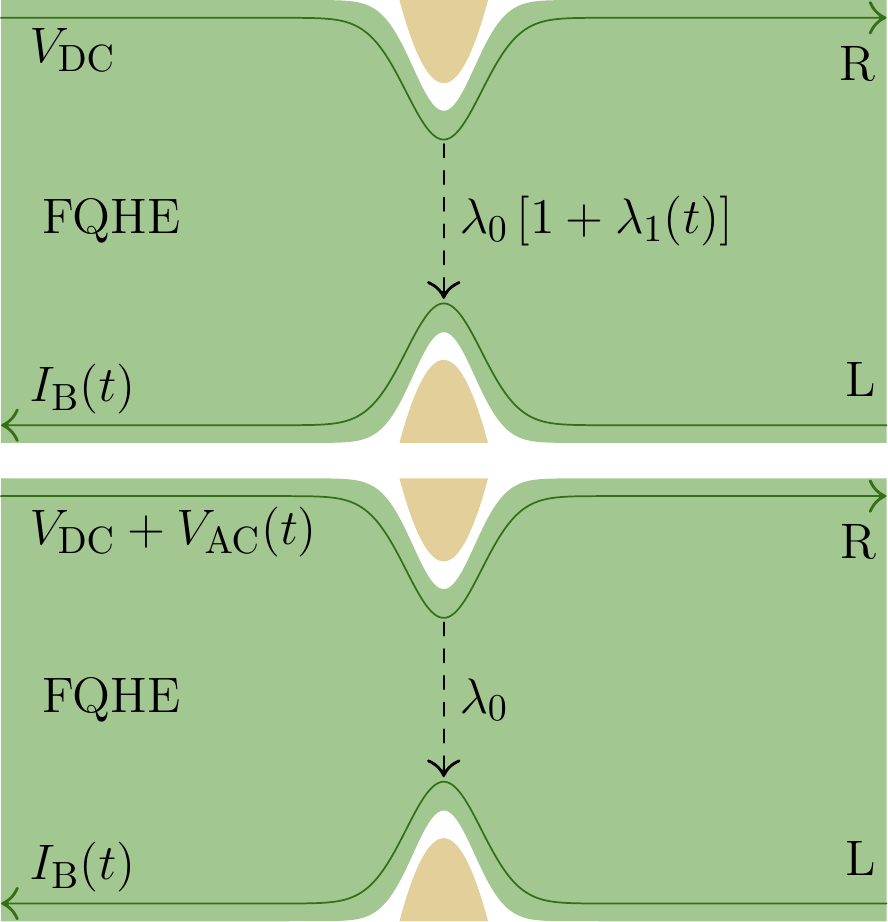}
	\caption{Sketch of the setup, a fractional quantum Hall bar pinched off by a quantum point contact. Top: AC gate drive setup where a DC voltage is applied between edge states and the tunnel coupling is time dependent. Bottom: AC voltage drive setup where a periodic voltage is applied on top of the DC voltage and the tunnel coupling is constant.}
	\label{fig:setup}
\end{figure}

The simplest quantum transport setup in the FQHE consists of a quantum Hall bar, along which edge excitations propagate, denoted left- and right-movers, further equipped with a QPC  (see Fig.~\ref{fig:setup}). Voltage sources can be connected to either edges in order to impose a potential bias difference between edge states, and the QPC can be tuned at will, with a special emphasis on two specific regimes. First, in the weak backscattering regime, the QPC is weakly pinched, the quantum Hall fluid spreads over the whole bar, and the dominant charge transfer process between the top and bottom edge is provided by quasiparticle excitations. In this situation, it is typically the backscattering current $ I_\text{T}$ which is computed/measured. This regime is depicted in both panels of Fig.~\ref{fig:setup}. Second, in the opposite limit, called the strong backscattering regime, the QPC is strongly pinched and the quantum Hall fluid is split in two (not shown). Only electrons can then tunnel between the left and right moving edges as they have to cross a vacuum region. The measured current corresponds then to that flowing between the left and right sides of the split Hall fluid.

Here, to obtain the scaling dimension of quasiparticles via a photo-assisted current measurement, we focus mainly on a weak backscattering situation, but results in the opposite regime of strong backscattering will also be presented for completeness. 

Assuming that a DC voltage $V_{\text{DC}}$ is imposed on the right-moving edge (see Fig.~\ref{fig:setup}), the tunnel, or backscattering Hamiltonian reads (see Ref.~\cite{rech17}):
\begin{equation}\label{eq:hamiltonian}
	H_\text{T} = \sum_{\epsilon = \pm} \left[ \lambda(t) e^{  i \omega_0^* t}  \psi_R^\dagger (0) \psi_L (0)  \right]^\epsilon\, ,
\end{equation}
where $\epsilon=+$ leaves the expression unchanged, while $\epsilon=-$ specifies the Hermitian conjugate. $\lambda(t)$ is the (time dependent, see below) tunnel coupling amplitude and $\omega_0^*=e^*V_{\text{DC}}$ in the weak backscattering regime. Below we consider two distinct setups for photo-assisted transport (upper and lower panels of Fig.~\ref{fig:setup}) which can both be described by a general form $\lambda(t)$ of the tunnel coupling. In full generality, it is assumed to be a (complex valued) periodic function of time. 

With these conventions, the backscattered current reads:
\begin{equation}
	 I_\text{T} (t) = i e^* \sum_{\epsilon = \pm} \epsilon \left[ \lambda(t) e^{i \omega_0^* t} \psi_R^\dagger (0,t) \psi_L (0,t)  \right]^\epsilon\, .
\end{equation}
We thus see that in the weak backscattering regime, the fractional charge $e^*=\nu e$ appears both as a prefactor of the current, and through the definition of the DC bias frequency $\omega_0^*$.

The strong backscattering equivalent of the tunnel Hamiltonian and current operator are achieved with the duality transformation $e^*\to e$, $\omega_0^*\to eV_{\text{DC}}$, $\psi \to \Psi$.

There are in fact two ways to achieve photo-assisted transport, both involving a constant gate voltage and a constant voltage bias $V_{\text{DC}}$.
1) one can apply an AC modulation on the QPC gate. This setup was discussed in Ref.~\cite{feldman03} in a very different context. 
2) one can directly superpose to the DC voltage drive an AC component. 
We call the former the gate drive and the latter the voltage drive, even though they both contain a DC voltage component.
This was studied theoretically and experimentally for superconducting hybrid junctions in Refs.~\cite{bertin-johannet22,schoelkopf98,kozhevnikov00}, and both theoretically and experimentally in Refs.~\cite{crepieux04,kapfer19,rech17,bisognin19} for a QPC in the FQHE regime. 
We note that the results of Refs.~\cite{lesovik94,lesovik99,torres01} for both normal metal junctions and normal metal/superconducting hybrid junctions fall in this category, although they do not directly consider a voltage drive.

Interestingly, the gate voltage modulation scenario has received little attention so far in the FQHE. 
In the core of this paper we wish to stress that it is especially relevant in the search for manifestations of the scaling dimension of quasiparticles. Both setups are depicted in Fig.~\ref{fig:setup}.

Note that both types of AC modulations can effectively be included in the tunnel amplitude. For the gate modulation, one adds an oscillating contribution $\lambda_1(t)$ to the bare tunnel amplitude $\lambda_0$, while for the voltage drive modulation, one incorporates the AC voltage drive via the Peierls substitution, i.e., as an additional phase, yielding
\begin{equation}\label{eq:lambdadef}
	\lambda(t)=\left\{\begin{aligned}
        	&\lambda_0(1+\lambda_1(t))\, &\text{gate drive,}\\
	        &\lambda_0\exp\left( i e^* \int_{-\infty}^t dt' V_{ac} (t') \right)\, &\text{voltage drive}\, ,
	\end{aligned}\right.
\end{equation}
where $\lambda=0$ corresponds to an infinite barrier. Concerning the gate drive, without loss of generality one can choose it to be real valued. It then only makes physical sense to choose $\vert \lambda_1(t)\vert <1$.

Assuming that both drives are periodic, and in order to stick to previous conventions \cite{bertin-johannet22}, we specify the Fourier decomposition of the tunnel coupling as:
\begin{equation}\label{eq:lambdafourier}
	\lambda(t) = \lambda_0\sum_l\overline{p_l}e^{il\Omega t}\, .
\end{equation}
where the drive frequency is $\Omega$, and $\overline{x}$ is the complex conjugate of $x$.
Then, provided that no assumptions are made on the $p_l$, the computation can be carried out simultaneously for the voltage drive or the gate drive. 
However, it is worth mentioning that the Fourier coefficients $p_l$ do not bear the same physical meaning when describing the two different drives. For the gate drive, there is typically a finite number of Fourier coefficients, which satisfy $\overline{p_l}=p_{-l}$ to represent a real valued modulation of the tunnel amplitude. However, for a voltage drive, the $p_l$'s are the Fourier coefficients of a complex number of modulus one. As a consequence, there is an infinite set of coefficients $p_l$ in this case, which obey specific sum rules \cite{dubois13b, bertin-johannet22}.

\section{Current for arbitrary \texorpdfstring{$\nu$}{nu}}\label{sec:currentcomp}

The aim of this section is to provide a general derivation of the backscattering current while making minimal assumptions on the Fourier coefficients $p_l$ of either drive. Also, specifically for this section, we keep in mind that at any moment, the duality transformation to the strong backscattering regime can be operated. Furthermore, we explicitly write the scaling dimension as $\nu_D$, although in the Laughlin case, it reduces to $\nu_D = \nu$. It allows to track down the effects which are specific to the scaling dimension. This choice is further reinforced by the general case detailed in Appendix~\ref{app:generalFQH}, where the degeneracy between scaling dimension and filling factor is lifted.

The photo-assisted backscattering current can be computed to second order in the tunnel coupling $\lambda(t)$, using the Keldysh formalism \cite{keldysh65}. 
From Eq.~\eqref{eq:quasiparticle_operator} and Ref.~\cite{martin05}, it reads
%\begin{widetext}
%\begin{equation}\label{eq:avgIB}
%	\langle I_\text{T} (t) \rangle =  
%	\frac{e^*}{2} \left( \frac{1}{2 \pi a} \right)^2 \sum_{\epsilon=\pm} \epsilon \int \mathrm{d}t' e^{i \epsilon \omega_0^* (t-t')} \left[ \lambda(t) \right]^{\epsilon} \left[ \lambda(t') \right]^{-\epsilon} \sum_{\eta,\eta'} \eta' e^{2 \nu_D \mathcal{G}^{\eta \eta'} \left( t-t' \right)} \, ,
%\end{equation}
%\end{widetext}
\begin{align}
	\langle I_\text{T} (t) \rangle =  &
	\frac{e^*}{2} \left( \frac{1}{2 \pi a} \right)^2 \sum_{\epsilon=\pm} \epsilon \int \mathrm{d}t' e^{i \epsilon \omega_0^* (t-t')} \nonumber \\
& \quad \times \left[ \lambda(t) \right]^{\epsilon} \left[ \lambda(t') \right]^{-\epsilon} \sum_{\eta,\eta'} \eta' e^{2 \nu_D \mathcal{G}^{\eta \eta'} \left( t-t' \right)} \, ,
\label{eq:avgIB}
\end{align}
where $\eta$, $\eta'$ are Keldysh contour indices and $\mathcal{G}^{\eta\eta'}$ is the corresponding bosonic Keldysh Green's function defined in Appendix \ref{app:keldyshgf}. %This expression illustrates that to this order in the tunneling amplitude, quasiparticle number is conserved on each edge. 
At this stage, it is important to stress out that this expression of the tunneling current readily generalizes to any Abelian edge theory comprising multiple bosonic modes, where it then depends on the effective charge and scaling dimension of the quasiparticle $\psi_{\mathbf{g}^*}$ involved in the leading tunneling process at the QPC, as we show in detail in Appendix~\ref{app:generalFQH},
\begin{widetext}
\begin{align}
\left\langle I_T (t) \right\rangle = \frac{1}{2} Q_{\mathbf{g}^*}   \sum_\epsilon \epsilon \int dt' 
e^{i \epsilon Q_{\mathbf{g}^*} V_\text{DC} (t-t')}
\left[  \Gamma_{\mathbf{g}^*} (t) \right]^\epsilon \left[\Gamma_{\mathbf{g}^*} (t') \right]^{-\epsilon} 
\sum_{\eta \eta'} \eta'  e^{2 \delta_{\mathbf{g}^*} \mathcal{G}^{\eta \eta'} (t-t')} ,
\end{align}
\end{widetext}
where $Q_{\mathbf{g}^*}$ and $\delta_{\mathbf{g}^*}$ are respectively the effective charge and scaling dimension of  the leading tunneling quasiparticle $\psi_{\mathbf{g}^*}$, while $\Gamma_{\mathbf{g}^*}$ is the corresponding tunneling amplitude. In particular, this expression underlines the importance of the distinction we put forward between scaling dimension and filling factor (as the two only turn out to be equal in the Laughlin case), and further emphasizes the major role played by the scaling dimension in our derivation.

The general calculation of the backscattering current at finite temperature and for arbitrary periodic drives is quite cumbersome, and details of the derivation are provided in Appendix~\ref{app:compsteps}. First, the summation over the Keldysh indices is performed explicitly using the symmetry properties of the chiral bosonic Green's function components. Next, the time integral is performed and written in terms of Gauss' hypergeometric function $_2F_1$. The result reads
\begin{widetext}

\begin{equation}\label{eq:currinter}
	\begin{aligned}
		\langle I_\text{T} (t) \rangle =  
		e^* &\left(2v \tau_0\right)^{-2}\pi^{-3}\beta\lambda_0^2\sum_{l,m}\overline{p_l}p_me^{i(l-m)\Omega t}\\
		\times\sum_{\eta=\pm}\eta&\Bigg[\frac{-i\eta\sin\left(\frac{\pi}{\beta}\tau_0\right)\exp\left(i\eta\frac{\pi}{\beta}\tau_0\right)}{\nu_D-i\frac{m+q}{2\pi\theta}}{_2F_1}\left(1, 1-\nu_D -i\frac{m+q}{2\pi\theta} ; 1+\nu_D-i\frac{m+q}{2\pi\theta}; \exp\left(2i\eta\frac{\pi}{\beta}\tau_0\right)\right)\\
		&\qquad\qquad\qquad\qquad\qquad\qquad\qquad\qquad\qquad\qquad\qquad\qquad\qquad\qquad\qquad\qquad-(m, q)\to(-l, -q)\Bigg]\, ,
	\end{aligned}
\end{equation}

\end{widetext}
where $\tau_0 = a/v$ is the short time cutoff, $\beta$ is the inverse temperature, $\theta=\left(\beta\Omega\right)^{-1}$ is the reduced temperature and $q=\omega_0^*/\Omega$.
$_2F_1$ is the Gauss hypergeometric function and the shorthand notation $f(a,b) - f(c, d) \equiv f(a,b)- (a, b)\to(c, d)$ has been used. An important advantage of this expression is that it remains valid for arbitrary values of the scaling dimension $\nu_D>0$, allowing us to also obtain the current in various limiting cases, including in the strong backscattering limit using the duality transformation. However, the convergence properties of the resulting hypergeometric function significantly depend on the value of the scaling dimension. For this reason, all physically relevant cases corresponding to, the Laughlin fractions for weak backscattering, the Fermi liquid case, or the Laughlin fractions for strong backscattering, have to be discussed separately. Note that this subtlety does not occur in the standard computation of the current in the presence of a DC voltage, it is specific to photo-assisted transport at finite temperature.

\section{Weak backscattering regime}\label{sec:weakbs}

Here, we focus on filling factors $0<\nu<1$ in Eq.~\eqref{eq:currinter}, corresponding solely to the weak backscattering regime dominated by quasiparticle transfer through the quantum Hall fluid. One can perform an expansion of the hypergeometric function ${_2F_1}$, which is specific to these filling factors, to leading order in $\tau_0/\beta$ (we recall that $\tau_0$ is the short time cutoff of the chiral Luttinger liquid theory). This is achieved in Appendix~\ref{app:wbcurrent}. Furthermore, without loss of generality (due a the choice of time origin), we can safely assume that the Fourier coefficients $p_l$ are real. This leads to the general expression for the current in terms of cosine and sine harmonics at the drive frequency:
\begin{widetext}
\begin{equation}\label{eq:current_cosine_sine_harmonics}
	\begin{aligned}
		\left\langle I_\text{T} (t) \right\rangle =I_0 + \mathcal{I} \sum_{l>0}\Bigg[&\cos(l\Omega t)\sum_{m}\left\lvert\Gamma\left(\nu_D+i\frac{m+q}{2\theta\pi}\right)\right\rvert^2(p_{m-l}p_m + p_m p_{l+m} )\sinh\left(\frac{m+q}{2\theta}\right)\\
		+&\sin(l\Omega t)\tan(\pi\nu_D)\sum_{m}\left\lvert\Gamma\left(\nu_D+i\frac{m+q}{2\theta\pi}\right)\right\rvert^2( p_m p_{l+m} - p_{m-l} p_m )\cosh\left(\frac{m+q}{2\theta}\right)\Bigg]\, ,
\end{aligned}
\end{equation}
\end{widetext}
with the prefactor:
\begin{equation}
	\mathcal{I}=\frac{e^*\Omega}{\pi}\left(\frac{\lambda_0}{ v}\right)^2\left(\frac{2\pi\theta}{\Lambda}\right)^{2\nu_D-2}\frac{\theta}{\Gamma(2\nu_D)}\,,
\end{equation}
where $\Lambda=(\Omega\tau_0)^{-1}$ is the reduced high energy cutoff, and the zeroth harmonic is
\begin{equation}
    I_{0} = \mathcal{I} \sum_{m} \left\lvert \Gamma \left(\nu_D+i\frac{m+q}{2\theta\pi}\right) \right\rvert^2 p_m^2 \sinh\left(\frac{m+q}{2\theta}\right) \, .
\label{eq:avgCurrent}
%\end{aligned}
\end{equation}

A few comments are in order at this stage. First, we stress that this formula for the fully time-dependent current is valid for both a voltage drive and a gate drive. 
Second, the zeroth harmonic contribution introduced in Eq.~\eqref{eq:avgCurrent} corresponds naturally to the current averaged over one period of the drive and, in the voltage drive case, satisfies a Tien-Gordon-like formula \cite{tien63} as it corresponds to a weighted sum of DC contributions with a shifted voltage $e^* V_\text{DC} \to e^* V_\text{DC} + m\Omega$ and probability $p_m^2$. Finally, while all harmonics of the current depend on the scaling dimension in a nontrivial way, it turns out that the sine harmonics, in $\sin(l\Omega t)$, all carry a prefactor $\tan(\pi\nu_D)$, which constitutes a striking dependence on the scaling dimension worth exploring further.  Note that a similar-looking dephasing in the time-dependent current has been obtained previously in some related cases \cite{feldman03, crepieux04} but remained unexploited. Indeed, it does not seem obvious to easily isolate this factor from the backscattering current, as the latter involves many contributions of the same order of magnitude.

To this end, the current can be rewritten as
\begin{equation}\label{eq:avIwb}
	\left\langle I_\text{T}(t) \right\rangle = \sum_{n=0}^\infty I_{n}(t)\, ,
\end{equation}
where
\begin{equation}
		I_{n\neq 0}(t)=\mathcal{I} C_n\cos(n\Omega t + \varphi_n)\, ,
\label{eq:Cnandphin}
\end{equation}
and the formulas for $\varphi_n$ and $C_n$ are given in Appendix.~\ref{app:wbcurrent}, see Eq.~\eqref{eq:fcoefwb}. 

We mention in passing that the analytical continuation of Eq.~\eqref{eq:current_cosine_sine_harmonics} for $\nu_D=1$ holds, allowing us to retrieve the Fermi liquid behavior discussed below (see Sec.~\ref{sec:fermiliquid}), although the computation steps are not quite valid in this regime.

\begin{figure*}
	\centering
	\includegraphics[width = 0.96\textwidth]{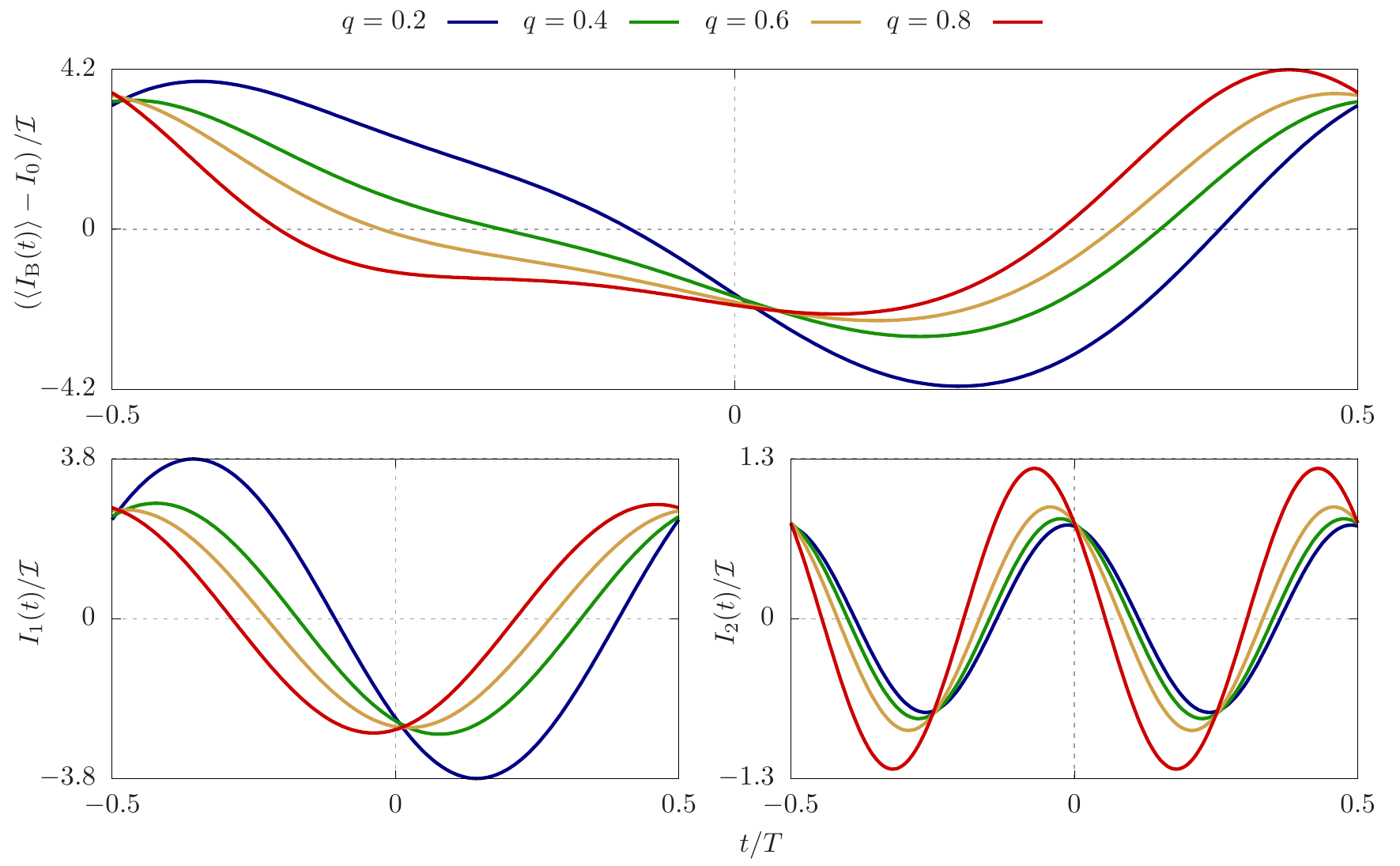}
	\caption{\textbf{Voltage drive case:} Average current through the QPC in the weak backscattering regime under the voltage drive $V(t) = V_{\text{DC}}+V_{\text{AC}}\cos{\Omega t}$ and for various values of the average transmitted charge $q = e^* V_{\text{DC}} / \Omega$. The filling factor is $\nu=1/3$, the AC normalized amplitude is $\alpha=1$, and the temperature is $\theta=0.1$. The top panel shows the AC part of the current as a function of time, over one period. The lower left panel displays the first harmonic and the lower right the second one.}
	\label{fig:wb-cont-cos}
\end{figure*}

\section{Harmonics of the current}\label{sec:harmonics}

This section is devoted to the analysis of the current and its different harmonics in the weak backscattering regime, as defined in Eq.~\eqref{eq:avIwb}.
We start by analyzing a cosine voltage drive, showing that there is no simple way to extract the scaling dimension of the quasiparticles from the current or its harmonics in this setting.
On the other hand, for a cosine gate drive, we establish, in a second subsection, a proportionality relation between the phase shift of the second harmonic of the current and the scaling dimension of the quasiparticle operator.

\subsection{Voltage drive}

When a voltage drive is applied, the tunnel coupling is modified according to Eq.~\eqref{eq:lambdadef}.
The drive is defined as
\begin{equation}\label{eq:voltagedrive}
	V(t) = V_{\text{DC}}+V_{\text{AC}}\cos{\Omega t}\, ,
\end{equation}
where the normalized modulation amplitude is $\alpha=e^*V_{\text{AC}}/\Omega$.
The Fourier coefficients of the tunnel coupling, see Eq.~\eqref{eq:lambdafourier}, read \cite{rech17}
\begin{equation} \label{eq:plBessel}
	p_l=J_l(-\alpha)\, ,
\end{equation}
where $J_l$ are Bessel functions of the first kind. One can readily check that these coefficients are real so that the expression for the time-dependent current, Eq.~\eqref{eq:current_cosine_sine_harmonics}, can be used as is.

It follows from Eq.~\eqref{eq:plBessel} that the Fourier coefficients $p_l$ are nonzero for any $l$. The harmonics $I_n$ of the current are therefore written as infinite sums, involving all Fourier coefficients. This significantly complicates the resulting expressions. As a result, in this voltage drive regime, we have been unable to extract a simple signature of the scaling dimension of the quasiparticle operator from the harmonics of the current. This situation is not specific of the present choice of a cosine drive, but instead arises from the time dependence of the tunnel coupling, which appears as an exponential of a periodic function.

An illustration of the fully time-dependent current is proposed in Fig.~\ref{fig:wb-cont-cos}. 
We show the AC part of the current and its first two harmonics as a function of time over a full period for various values of the reduced DC voltage $q<1$. We remark that the current displays a rich behavior, in particular, both the amplitude and the phase of the two first harmonics of the current depend on $q$ in a nontrivial way. Indeed, each harmonic involves a large number of Fourier components of the tunnel coupling, making it impractical to extract any valuable information.  
%This situation is therefore not useful to extract the scaling dimension of the quasiparticle operator.

\subsection{Gate drive}

The tunnel coupling under a gate drive, as defined in Eq.~\eqref{eq:lambdadef}, reads, for a cosine drive, 
\begin{equation}\label{eq:gatedrive}
	\lambda(t)=\lambda_0 \left[ 1+\lambda_1\cos(\Omega t) \right] \, .
\end{equation}
Therefore, its Fourier coefficients are given by
\begin{equation}
		p_0=1\, ,\qquad p_{\pm 1} = \frac{\lambda_1}{2}\, ,\qquad p_{\lvert n\rvert>1}=0\, .
\end{equation}
These coefficients are real, allowing us to use the expression for the fully time-dependent current of Eq.~\eqref{eq:current_cosine_sine_harmonics}. More importantly, there is only a finite subset of coefficients that are nonzero (two in the present case). This is a major difference between the gate drive and the voltage drive. While in the latter case, the proliferation of nonzero coefficients did not allow us to obtain a simple self-contained expression of the current, in the present case of a gate drive, the internal summations over $l$ and $m$ in Eq.~\eqref{eq:current_cosine_sine_harmonics} can be readily performed. The resulting expression for the fully time-dependent current is still quite cumbersome. However, working out explicitly the expression for the amplitudes $C_n$ and phases $\varphi_n$ [see Eq.~\eqref{eq:Cnandphin}], one can show that
\begin{widetext}
\begin{align}
\tan \varphi_2 = \tan \pi \nu_D 
\frac{
\left\lvert\Gamma\left(\nu_D+i\frac{q+1}{2\theta\pi}\right)\right\rvert^2 \cosh\left(\frac{q+1}{2\theta}\right)
-
\left\lvert\Gamma\left(\nu_D+i\frac{q-1}{2\theta\pi}\right)\right\rvert^2 \cosh\left(\frac{q-1}{2\theta}\right)
}{
\left\lvert\Gamma\left(\nu_D+i\frac{q+1}{2\theta\pi}\right)\right\rvert^2 \sinh\left(\frac{q+1}{2\theta}\right)
+
\left\lvert\Gamma\left(\nu_D+i\frac{q-1}{2\theta\pi}\right)\right\rvert^2 \sinh\left(\frac{q-1}{2\theta}\right)
}
\label{eq:tanphi2}
\end{align}
\end{widetext}
which, in the low temperature regime ($\theta \ll 1$), further reduces to
\begin{align}
\varphi_2 = \pi \nu_D 
\end{align}
provided that the reduced DC voltage satisfies $q < 1$. Quite astonishingly, the phase shift of the second harmonic of the current induced by a cosine gate drive in the low temperature limit is exactly equal to the scaling dimension of the quasiparticle operator.
%We stress one more time that this result relies deeply on the fact that there are only few Fourier coefficients which are linear in the tunnel amplitude $\lambda_1$.

\begin{figure*}
	\centering
	\includegraphics[width=0.96\textwidth]{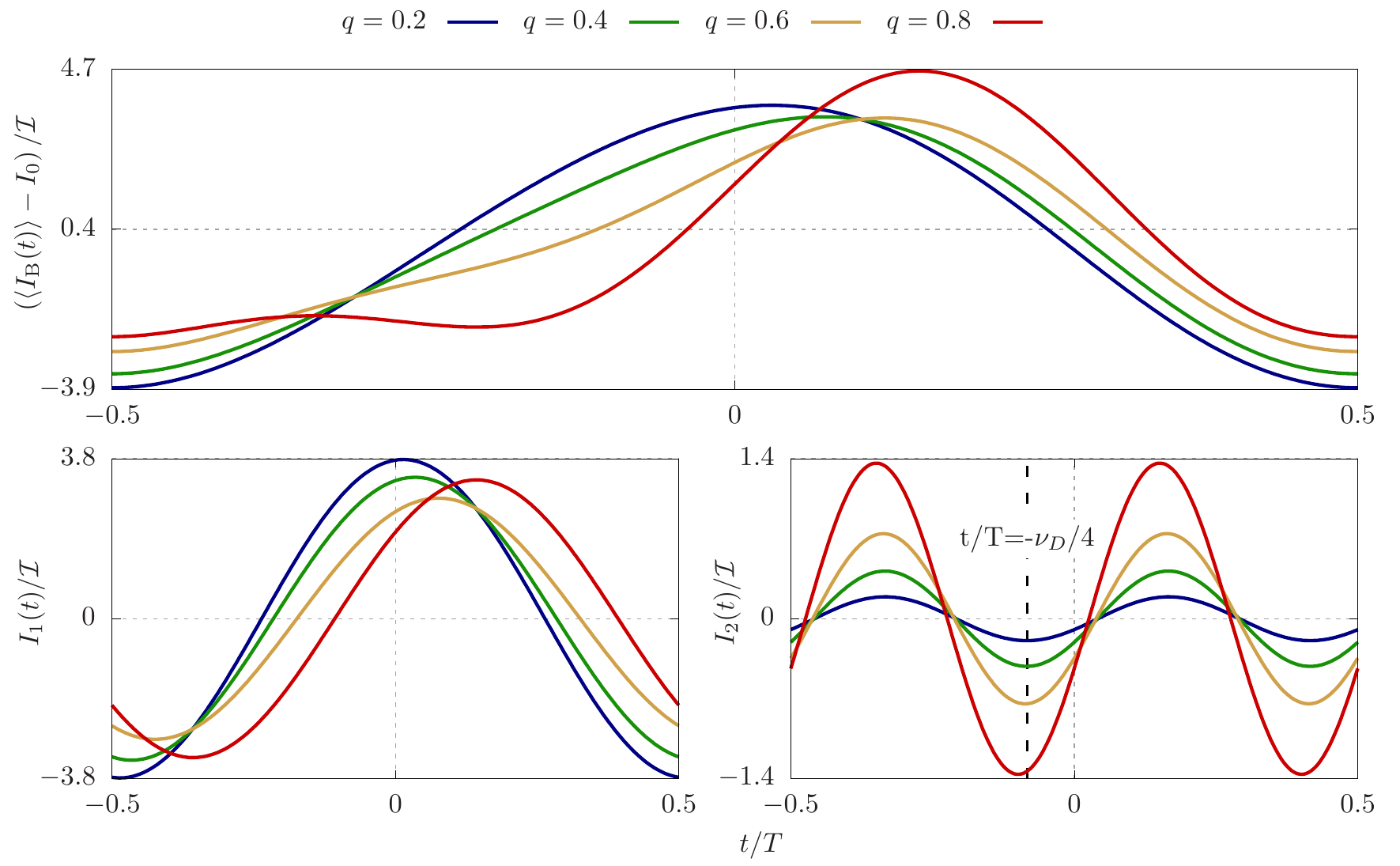}
	\caption{\textbf{Gate drive case:} Average current through the QPC in the weak backscattering regime under the gate drive Eq.~\eqref{eq:gatedrive} and for various values of $q$. The figure was obtained with the following parameters: the filling factor is $\nu=1/3$, the tunneling amplitude modulation is $\lambda_1=1$ and the reduced temperature is $\theta=0.1$. The top panel shows the AC part of the current as a function of time, over one period. The lower left panel displays the first harmonic of the current and the lower right panel the second harmonic of the current, where all curves depict the same phase shift set by the scaling dimension $\nu_D$.}
	\label{fig:wb-gate-cos}
\end{figure*}

The same representation of the current as that adopted for the voltage drive  (see Fig.~\ref{fig:wb-cont-cos}) is proposed in Fig.~\ref{fig:wb-gate-cos}, for a finite reduced temperature $\theta=0.1$.
In the lower right panel we remark that the phase shift of the second harmonic is indeed equal to $\pi\nu_D$ for a large range of $q$ (the slight discrepancy for the highest $q=0.8$ disappears at lower temperature). Indeed, from Eq.~\eqref{eq:Cnandphin}, one readily sees that this phase shift can be recast as a shift in time by an amount $t/T = - \varphi_2 / (4\pi) = - \nu_D/4$ (taking into account that $C_2 > 0$). 

\begin{figure}
	\centering
	\includegraphics[width = 0.48\textwidth]{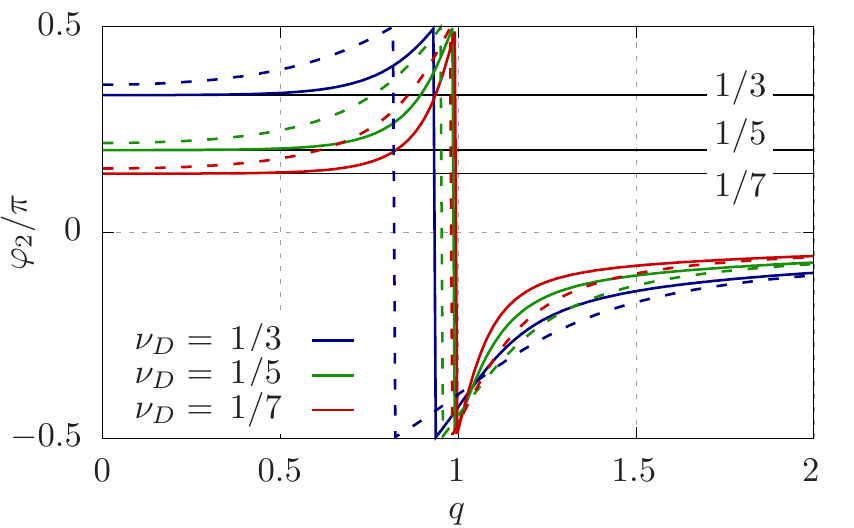}
	\caption{Phase of the second harmonic [as defined in Eqs.~\eqref{eq:fcoefwb} and \eqref{eq:tanphi2}] of the current through the QPC under the gate drive Eq.~\eqref{eq:gatedrive}, in the weak backscattering regime, as a function of $q$, for various filling factors $\nu$. Full lines are for reduced temperature $\theta=0.1$ and dashed lines for $\theta=0.2$.}
	\label{fig:wb-gate-cos-phase}
\end{figure}

The robustness of this result for finite temperature is explored in Fig.~\ref{fig:wb-gate-cos-phase}. It displays the evolution of $\varphi_2$ as a function of $q$ for different values of $\nu_D$ and at two experimentally realistic reduced temperatures, $\theta=0.1$ and $\theta=0.2$. Note that in actual experimental realizations, a reduced temperature $\theta=0.1$ would correspond to an actual temperature of $50\milli\kelvin$ for a drive frequency of $10\giga\hertz$. For this value of the reduced temperature, the results of Fig.~\ref{fig:wb-gate-cos-phase} show a good agreement between the phase shift $\varphi_2$ and the scaling dimension, over a large range of DC voltage. Increasing the temperature leads to a small departure between the two, which further grows as one increases the reduced voltage $q$ (as already observed for $q=0.8$ in Fig.~\ref{fig:wb-gate-cos}).

We stress that the identification of the phase shift, which gives direct access to the quasiparticle operator scaling dimension requires the experimental measurement of the {\it time-dependent current}, rather than the measurement of its average value over the period of the drive. Experimentally, it would therefore be necessary to measure the {\it harmonics} of the current, for instance by multiplying the current signal by a chosen, specific, periodic signal, and subsequently performing the average over the period of the drive. 

The present prediction for the phase shift as a signature of the scaling dimension of the quasiparticle operator for a gate voltage modulation constitutes the central result of this work. Based on the generalized derivation presented in Appendix~\ref{app:generalFQH}, and the assumptions underlying the above computations, it should hold for a broad range of filling factors, the only requirement being that the scaling dimension of the quasiparticle involved in the leading tunneling process satisfies $0 < \nu_D < 1$ with $\nu_D \neq 1/2$.

\section{Fermi liquid computation}\label{sec:fermiliquid}

In this section we propose a computation of the Fermi liquid limit, first, using standard Fermi liquid theory and second, taking the limit $\nu=1$ of the chiral Luttinger liquid theory presented above.
This fulfills two purposes, it extends to periodic drives the usual Fermi liquid computation of the current through a QPC and it allows to perform consistency checks of the Fermi liquid limit of our Luttinger liquid computation.

\subsection{Fermi liquid formalism}
In the Fermi liquid picture, the Hamiltonian is written in analogy with Eq.~\eqref{eq:hamiltonian}, in terms of electron creation and annihilation operators $\Psi_{\text{L,R}}$ at position $x=0$ in the left or right leads.
It reads
\begin{equation}
    \mathcal{H}_\text{T}=\lambda(t)\Psi_\text{L}^\dagger \Psi_\text{R}+\text{H.c.}
\end{equation}
Thus the current operator is
\begin{equation}
    I_\text{T}=ie\lambda(t)\Psi_\text{L}^\dagger \Psi_\text{R}+\text{H.c.}
\end{equation}
We define Keldysh Green's functions for electron operators as
\begin{equation}
    G_{\text{ss'}}^{\eta\eta'}=-i\left\langle \mathcal{T}_K\Psi_\text{s}(t_\eta)\Psi_\text{s'}^\dagger(t'_{\eta'}) \right\rangle\, ,
\end{equation}
where $\mathcal{T}_{\text{K}}$ is the time ordering operator along the Keldysh contour, $\text{s}$ and $\text{s'}$ can be either $\text{L}$ or $\text{R}$ and the Keldysh contour indices $\eta$ and $\eta'$ can be $+$ or $-$.
The average current can be written with Keldysh Green's function for electron operators:
\begin{equation}
    \left\langle I_\text{T}(t)\right\rangle = -e\left[\lambda(t)G_{\text{RL}}^{+-}(t,t) - \lambda^*(t)G_{\text{LR}}^{+-}(t,t)\right]\, .
\end{equation}

Working in the wide band limit, the expression for the current can be obtained using standard tools (see Appendix \ref{app:fermiliq} for details) and can therefore be simplified into
\begin{align}
    \left\langle I_\text{T}(t)\right\rangle =& \frac{\lambda_0^2 e}{4\pi v_{\text{F}}^2}\sum_{l,m} \overline{p_l}p_me^{i(l-m) \Omega t}
\nonumber \\
& \times \int\mathrm{d}\omega
\left[ f(\omega-\mu_\text{R}) - f(\omega+l\Omega-\mu_\text{L}) \right. \nonumber \\
& \qquad \left. -f(\omega-\mu_\text{L}) + f(\omega+m\Omega-\mu_\text{R})\right]\, .
\end{align}
where $f(x)$ is the usual Fermi distribution and $\mu_{R/L}$ is the chemical potential of edge $R/L$. Performing the integration yields
\begin{equation}\label{eq:flresult}
    \left\langle I_\text{T}(t)\right\rangle = \frac{e\Omega\lambda_0^2}{4\pi v_F^2}\sum_{l,m} \overline{p_l}p_me^{i(l-m) \Omega t}(2q+l+m)\, .
\end{equation}
Using Eq.~\eqref{eq:lambdafourier} to write the derivative
$\partial_t \lambda(t) = i\Omega\sum_l l\overline{p_l} e^{il\Omega t}$,
the current can be rewritten as
\begin{equation}\label{eq:flinter}
	\langle I_\text{T} (t) \rangle = 
\frac{e\Omega}{4\pi v_{\text{F}}^2}
\left[
2q \left\lvert\lambda(t)\right\rvert^2 + \frac{1}{i\Omega} \partial_t \left\lvert\lambda(t)\right\rvert^2
\right]
\, .
\end{equation}

Finally, substituting the expression for the time-dependent tunnel coupling $\lambda (t)$ for the two types of drive, one has
\begin{equation}\label{eq:aclandauer}
    	\langle I_\text{T} (t) \rangle = \frac{e^2 }{2\pi v_{\text{F}}^2}\left\{{\begin{aligned}
			\lambda(t)^2V_{\text{DC}}&\qquad\text{for a gate drive}\\
			\lambda_0^2V(t)&\qquad\text{for a voltage drive}
		\end{aligned}}\right.
\end{equation}
Which is the straightforward time-dependent generalization of the Landauer formula for PAT.
The absence of a temperature dependence is a consequence of the wide band approximation.

\subsection{Luttinger liquid approach at \texorpdfstring{$\nu=1$}{nu=1}}

Finding the current in the Fermi liquid limit of the Luttinger theory can be done by setting $\nu_D=\nu=1$ in the general formula for the current, Eq.~\eqref{eq:currinter}. 
In this case, the expansion of Eq.~\eqref{eq:currinter} performed in Sec.~\ref{sec:weakbs} does not hold. 
However, as shown in Appendix \ref{app:sbcurrent}, a logarithmic expansion can be carried out to lowest order in $\tau_0/\beta$.
This expansion yields two sums, which we call Fermi/DC [Eq.~\eqref{eq:dcfermi}] and correlated-AC [Eq.~\eqref{eq:accor}], for reasons that will become clear below.
In the particular case of $\nu_D=1$, the leading term is the first term of the Fermi/DC sum, see Eq.~\eqref{eq:luttingerfermiliquid}, and the current is identical to that 
obtained from the Fermi liquid approach, Eq.~\eqref{eq:flresult}. We have thus checked the consistency of our chiral Luttinger liquid approach in the Fermi liquid regime.

\section{Strong backscattering regime}\label{sec:strongbs}

In this section we describe the behavior of the current in the strong backscattering regime, which is obtained from a duality transformation. The latter only holds for filling factors in the Laughlin series, so that, for clarity, we revert to a description in terms of the filling factor $\nu$. This regime is obtained by setting $\nu \to \nu^{-1}$, $e^*\to e$ and $\omega_0^*\to eV_{\text{DC}}/\Omega$ in the expression for the current, Eq.~\eqref{eq:avgIB}.
As $\nu^{-1}\in\mathds{N}$ for Laughlin fractions, one has to exploit the expansion of Eq.~\eqref{eq:currinter} to leading order in the cutoff, which is valid for all integers $\nu^{-1}>0$, i.e. the logarithmic expansion, Eq.~\eqref{eq:logexpansion}.

As already pointed out in Section~\ref{sec:fermiliquid}, this expansion consists of two sums.
In the case $\nu=1$, the leading term in $\tau_0/\beta$ belongs to the Fermi/DC sum.
However, here, the expansion in orders of $\tau_0/\beta$ favors another term, which belongs to the second sum, which we call correlated-AC sum, see Eq.~\eqref{eq:accor}. 
More precisely, the leading term is of first order in $\tau_0/\beta$ and yields a current:
\begin{equation}\label{eq:currentsbfinal}
\begin{aligned}
		\langle I_\text{T} (t) \rangle\approx&\frac{-e}{(1-2\nu^{-1})_3}\left(\frac{\lambda_0}{\pi v}\right)^2\frac{\Omega}{\Lambda}\\
		&\times\Im\left[\sum_{l,m}\overline{p_l}p_me^{i(l-m)\Omega t}\left(m+q\right)^2\right]\, ,
\end{aligned}
\end{equation}
where $(x)_n$ is the Pochhammer symbol, defined in Appendix~\ref{app:sbcurrent}. 
This expression can take a simpler form, since following the steps used to obtain Eq.~\eqref{eq:flinter}, one can write
\begin{widetext}
\begin{equation}
    \langle I_\text{T} (t) \rangle = \frac{e}{(2\nu^{-1}-1) (2\nu^{-1}-2) (2\nu^{-1}-3)} \frac{1}{\pi^2 v^2 \Omega \Lambda}\Im\left[\lambda_0^2e^2V_{\text{DC}}^2 + 2 i eV_{\text{DC}}\lambda(t)\partial_t\overline{\lambda(t)} - \lambda(t)\partial_t^2\overline{\lambda(t)}\right]\, .
\end{equation}
\end{widetext}

Finally, computing explicitly the imaginary part, one can express the end result in a unified way for both types of drives defined in Eq.~\eqref{eq:lambdadef} as
\begin{equation}\label{eq:derivaclandauer}
    \langle I_\text{T} (t) \rangle = \frac{e}{\left( \frac{2}{\nu}-1 \right) \left( \frac{2}{\nu}-2 \right) \left( \frac{2}{\nu}-3 \right)} \frac{1}{\pi^2 v^2 \Omega \Lambda} \partial_t\left(\lambda^2V\right)\, .
\end{equation}
%\begin{equation}\label{eq:derivaclandauer}
%    	\langle I_\text{T} (t) \rangle =   -\frac{2e^2}{\nu_D^{-1} \left(4\nu_D^{-2}-1 \right) v^{2}\Omega\Lambda}   \left\{{\begin{aligned}
%			V_{\text{DC}} \partial_t \lambda(t)^2 &\qquad\text{for a gate drive}\\
%			\lambda_0^2  \partial_t V(t)&\qquad\text{for a voltage drive}
%		\end{aligned}}\right.
%\end{equation}
This result is quite intriguing as it involves the time derivative of the AC Landauer formula, Eq.~\eqref{eq:aclandauer}.
As in the Fermi liquid computation, there is no temperature dependence to this order, this reflects the wide band limit of the Luttinger model.

In the case of a voltage drive, the junction in the strong backscattering regime behaves as a standard capacitor, i.e.,
\begin{equation}
    \langle I_\text{T} (t) \rangle=C\frac{\mathrm{d}V(t)}{\mathrm{d}t}\, ,
\end{equation}
with a capacitance $C=\frac{e^2}{\left( \frac{2}{\nu}-1 \right) \left( \frac{2}{\nu}-2 \right) \left( \frac{2}{\nu}-3 \right)} \frac{\lambda_0^2}{\pi^2 v^2 \Omega \Lambda}$, which, after restoring the proper powers of $\hbar$, can be further recast as $C= c \frac{2 \pi a}{\left( \frac{2}{\nu}-1 \right) \left( \frac{2}{\nu}-2 \right) \left( \frac{2}{\nu}-3 \right)} \left( \frac{\lambda_0}{\hbar v}\right)^2$, where $c= e^2/(h v)$ is the quantum capacitance by unit length.

In the case of a gate drive the situation is different, defining the transmission of the junction as $\tau(t)=4\lambda^2(t)$, the current reads
\begin{equation}
    \langle I_\text{T} (t) \rangle = \frac{e^2V_\text{DC}}{\left( \frac{2}{\nu}-1 \right) \left( \frac{2}{\nu}-2 \right) \left( \frac{2}{\nu}-3 \right) \pi^2 v^{2}\Omega\Lambda}\frac{\mathrm{d}\tau(t)}{\mathrm{d}t} \, .
\end{equation}

%As we find very different solutions from the same initial formula Eq.~(\ref{eq:avgIBgen}) at integer $\nu$, but in different regimes, we feel that a reminder of the differences between these regimes might prove helpful.

To summarize, the general expansion of the hypergeometric function in $\tau_0/\beta$ for positive integer $\nu_D^{-1}$ yields a current consisting of two sums, see Eq.~\eqref{eq:logexpansion}. 
We have to consider three different situations. 
When the junction is driven by a DC drive only, when the junction is driven by an AC drive and it is in the Fermi regime, $\nu=1$, or when the junction is driven by an AC drive and correlations are present, i.e., $\nu^{-1}>1$.

When the junction is in the Fermi liquid regime ($\nu=1$) or solely driven by a DC drive, the leading terms of the expansion belong to the same sum which we therefore call the Fermi/DC sum, see Eq.~\eqref{eq:dcfermi}.
In the DC case, this term yields the duality transformation of the already known weak backscattering DC result, see \cite{rech17}.
In the Fermi liquid case we find a straightforward extension of the Landauer formula which we call the AC Landauer formula.

When the junction is driven by both a DC and an AC drive (applied to either the edge or the gate) and correlations are present, i.e. $\nu^{-1}>1$, the leading contribution to the current comes from another term, which we therefore call the correlated-AC sum, Eq.~\eqref{eq:accor}. This term yields a current proportional to the time derivative of the AC Landauer formula, Eq.~\eqref{eq:derivaclandauer}.

%\section{Strong backscattering regime}\label{sec:strongbs}
%A few things to say, maybe first the fact that the DC current is subleading, the AC current is orders of magnitude higher and does not depend on the temperature (experimentally interesting ??). Also this average current is proportional to the derivative of the applied voltage.
%Second thing is that there is a dephasing of $\pi/2$ which I do not understand, is it due to the capacitance behavior ?

\section{Conclusion}\label{sec:conclusion}

In the introduction of this paper, we have stressed the considerable theoretical investment for the search of signatures  of the statistics of anyons of the FQHE in the context of electronic quantum transport setups. Most of these setups indeed require quite complicated geometries or types of measurements. Noting that the statistical angle of anyonic quasiparticles is intimately tied to the scaling dimension of the quasiparticle operator, we have asked a naive question: can this scaling dimension be detected via a careful measurement of the time-dependent backscattering current in the weak backscattering regime? 

For this purpose, we have reexamined the theory of photo-assisted transport in the FQHE. We have noted that such PAT can be achieved in two distinct ways. Either one modulates the gate voltage applied to the QPC (to our knowledge this type of drive has not received much attention in the context of PAT), or one adds an AC modulation on top of the DC voltage drive (as was proposed theoretically in Refs. \cite{crepieux04,rech17}). 

Our first task was to show that both drives can be described with a unified approach: the only difference between the two drives resides in the details of the Fourier decomposition of the tunnel amplitude describing them. The time-dependent current can then be computed analytically in terms of sums over these Fourier coefficients, further involving a Gauss hypergeometric function, a result which is quite abstract in nature. In order to make progress, expansions to leading order in the cutoff need to be subsequently performed. Interestingly, these expansions depend crucially on the value of the scaling dimension $\nu_D$ and whether the QPC is in the weak or strong backscattering regime.

For the weak backscattering regime, we obtained expressions which allow to characterize the time dependent current as a constant term accompanied by its harmonics at multiples of the drive frequency. It is precisely in these harmonics that we believe that it is possible to isolate the scaling dimension of the quasiparticle tunneling operator. Indeed, by choosing a simple cosine modulation for a gate drive, we were able to show that the phase shift of the second harmonic of the time-dependent current is directly proportional to the scaling dimension at low temperature, which constitutes the central message of this paper. We stressed that this connection is robust at finite temperature, rendering it accessible to experimental observation. A typical experiment would require to multiply the time dependent current signal by a proper harmonic drive at twice the drive frequency to detect this phase shift. 

For completeness, we further explored the expansion properties of hypergeometric functions which appear in the general expression of the time dependent current, in order to derive results for both the Fermi liquid limit $\nu = 1$ and for the strong backscattering limit. In the former case, we performed an independent Fermi liquid calculation using the Dyson equation for fermionic Keldysh Green's function and derived a time dependent generalization of the Landauer formula. We further showed that this AC Landauer formula is in agreement with the $\nu=1$ limit of the chiral Luttinger model. In the case of strong backscattering, we derived an expression of the current in terms of time derivative of the drives. 

Our main result about the phase shift suggest that a careful measurement of the time-dependent current for a device containing a {\it single} quantum point contact could provide an insight on the detection (albeit indirect) of fractional statistics in the fractional quantum Hall effect. 

\begin{acknowledgments}
Early discussions on gate modulation in the FQHE with Adeline Crépieux are acknowledged. We are also grateful to G. F\`eve and M. Hashisaka for useful discussions. 
This work received support from the French government under the France 2030 investment plan, as part of the Initiative d'Excellence d'Aix-Marseille Université - A*MIDEX.
We acknowledge support from the institutes IPhU (AMX-19-IET-008) and AMUtech (AMX-19-IET-01X). D.C.G. acknowledges the ANR FullyQuantum 16-CE30-0015-01 grant and the H2020 FET-OPEN UltraFastNano No. 862683 grant.
\end{acknowledgments}

\onecolumngrid

%%%%%%%%%%%%%%%%%%%%%%%%%%%%%%%%%%%%%%%%%
% appendix A
\begin{appendix}
	\section{Keldysh Green's functions relations}\label{app:keldyshgf}
	In this appendix we introduce the bosonic Keldysh Green's functions used in the main text, see Eq.~\eqref{eq:avgIB}. 
	They are defined as
\begin{equation}
	\mathcal{G}^{\eta \eta'}(t-t')=\left\langle\ \mathcal{T}_K \phi\left(t_\eta, x=0\right)\phi\left(t'_{\eta'}, x=0\right)\right\rangle\, ,
	\label{chiral_green}
\end{equation}
where $\mathcal{T}_K$ denotes ordering along the Keldysh contour (see Ref.~\cite{keldysh65}).
The four different Keldysh Green's functions can be summarized by a single one (which has no Keldysh indices)
\begin{equation}\label{eq:GFdef}
	\begin{aligned}
		\mathcal{G}^{++}(t-t')&=\mathcal{G}(\lvert t-t'\rvert)\\
		\mathcal{G}^{--}(t-t')&=\mathcal{G}(-\lvert t-t'\rvert)\\
		\mathcal{G}^{+-}(t-t')&=\mathcal{G}(t'-t)\\
		\mathcal{G}^{-+}(t-t')&=\mathcal{G}(t-t')\, ,
    \end{aligned} 
\end{equation}
where the `modified` Green's function is defined as
\begin{equation}\begin{aligned}
	\mathcal{G}(t-t')&=\left\langle\phi_{R(L)}(t)\phi_{R(L)}(t')\right\rangle\\
	&-\left\langle\phi_{R(L)}(t)^2\right\rangle/2 - \left\langle\phi_{R(L)}(t')^2\right\rangle/2\, ,
	\end{aligned}
\end{equation}
and reads \cite{martin05}
\begin{equation}\label{eq:kgf}
	\mathcal{G}(\tau) = - \log \left[ \frac{\sinh \left( \frac{\pi}{\beta} (i \tau_0 - \tau) \right) }{ \sinh \left( i \frac{\pi}{\beta} \tau_0 \right) } \right]\, ,
\end{equation}
where $\beta$ is the inverse temperature and $\tau_0\equiv a/v$ is the short time cutoff of the chiral Luttinger model.

%%%%%%%%%%%%%%%%%%%%%%%%%%%%%%%%%%%%%%%%%%%%%%%%%%%%%%%
%    appendix arbitrary nu

\section{General model for Abelian fractional quantum Hall edge states}\label{app:generalFQH}

In this appendix, we consider a general model for Abelian fractional quantum Hall edge states and show that it leads to a tunneling current of the same form as the one obtained in Eq.~\eqref{eq:avgIB}.

Our starting point is the action of the general Abelian FQH edge
\begin{align}
S = \frac{1}{4 \pi} \int dx dt \sum_{l=1}^N \left[ - \chi_l \partial_x \phi_l \partial_t \phi_l - v_l \left( \partial_x \phi_l \right)^2 \right] ,
\end{align}
where $\phi_l$ are a set of bosonic modes (with $l=1,...,N$), with chirality $\chi_l = \pm 1$ and velocity $v_l$. In the case of a single mode ($N=1$) this reduces to the standard description of the Laughlin series.

These fields satisfy the commutation relation
\begin{align}
\left[ \phi_l (x,t) , \phi_{l'} (x', t') \right] =i \pi \chi_l \delta_{ll'} \text{Sgn} \left( x-x' - \chi_l v_l t + \chi_{l'} v_{l'} t' \right) ,
\end{align}
and help defining the density operator as
\begin{align}
\rho = \frac{1}{2 \pi} \sum_l q_l \partial_x \phi_l ,
\end{align}
where the set of coefficients $q_l$ encode the contribution of the $l^\text{th}$ mode to charge transport. These coefficients are related to the filling factor in a nontrivial way as they satisfy the sum rule
\begin{align}
\sum_l \chi_l q_l^2 = \nu .
\end{align}

In analogy with the Laughlin case, the edge supports quasiparticles, whose creation/annihilation operators involve a linear combination of all bosonic modes, namely
\begin{align}
\psi_\mathbf{g} (x,t) \propto \exp \left[ i \sum_{l=1}^N g_l \phi_l (x,t) \right] .
\end{align}
For a given vector $\mathbf{g} = \left\{ g_1, ..., g_N \right\}$, the corresponding quasiparticle $\psi_\mathbf{g}$ is characterized by 3 important physical quantities:
\begin{align}
Q_\mathbf{g} &= e \sum_l \chi_l q_l g_l  &  \text{its effective charge} \\
\delta_\mathbf{g} &= \sum_l g_l^2 &  \text{its scaling dimension} \\
\Theta_\mathbf{g} &= \pi \sum_l \chi_l g_l^2 & \text{its statistical angle}
\end{align}
Note that in all generality, the statistical angle is bounded by the scaling dimension, $|\Theta_\mathbf{g}| \leq \pi \delta_\mathbf{g}$, and even reduces to $|\Theta_\mathbf{g}| = \pi \delta_\mathbf{g}$ in the special situation where all modes have the same chirality, $\chi_l = \chi$, $\forall l$.

The QPC is set in the weak backscattering regime and is thus modeled by a Hamiltonian describing the tunneling of quasiparticles between the two edges as
\begin{align}
H_T = \sum_\mathbf{g} \Gamma_\mathbf{g} {\psi_\mathbf{g}^{(u)}}^\dagger (0) \psi_\mathbf{g}^{(d)} (0) + \text{H.c.}
\end{align}
where $(u)/(d)$ label the upper and lower edges (the standard $R/L$ designation being ill-defined in the presence of non-chiral modes). In all generality, one would need to account for all possible tunneling events, i.e. ones involving all possible quasiparticles. In practice, however, it makes sense to favor the one with the lowest scaling dimension, as it is the most relevant perturbations in the RG sense. In what follows, we label this leading quasiparticle with the vector $\mathbf{g}^*$.

From the expression of the tunneling Hamiltonian, one readily obtains the tunneling current operator at the location of the QPC, as
\begin{align}
I_T (t) = i Q_{\mathbf{g}^*} \left[ \Gamma_{\mathbf{g}^*} (t) e^{i Q_{\mathbf{g}^*} V_\text{DC} t} {\psi_{\mathbf{g}^*}^{(u)}}^\dagger (0,t) \psi_{\mathbf{g}^*}^{(d)} (0,t) - \text{H.c.} \right] ,
\end{align}
where we introduce the effect of an applied DC voltage between edges and introduced a time-dependent tunnel coupling along the same lines as we did in the text, leading to Eq.~\eqref{eq:lambdadef}.

Using the decomposition of the quasiparticle operators in terms of the bosonic fields $\phi_l$, one can express the thermal average of the tunneling current in terms of the bosonic Green's function $\mathcal{G}_l^{\eta \eta'} (t-t')$ yielding
\begin{align}
\left\langle I_T (t) \right\rangle = \frac{1}{2} Q_{\mathbf{g}^*} \sum_{\eta \eta'} \eta' \int dt' 
\left[ 
\Gamma_{\mathbf{g}^*} (t) \overline{\Gamma_{\mathbf{g}^*} (t')} e^{i Q_{\mathbf{g}^*} V_\text{DC} (t-t')}
-
\overline{\Gamma_{\mathbf{g}^*} (t)} \Gamma_{\mathbf{g}^*} (t')  e^{i Q_{\mathbf{g}^*} V_\text{DC} (t-t')}
\right] \prod_l e^{2 {g_l^*}^2 \mathcal{G}_l^{\eta \eta'} (t-t')} .
\label{eq:interIT}
\end{align}
At this stage, it is important to keep in mind that $\mathcal{G}_l^{\eta \eta'} (t-t')$  is a trivial generalization of the one presented in Appendix \ref{app:keldyshgf}. In particular, its Keldysh components follow the same relations as the ones introduced in Eq.~\eqref{eq:GFdef}, with the corresponding Green's function $\mathcal{G}_l (\tau)$ given by
\begin{align}
\mathcal{G}_l (\tau) = - \log \left[ \frac{\sinh \left( \frac{\pi}{\beta} \left(i \tau_l - \tau\right) \right) }{ \sinh \left( i \frac{\pi}{\beta} \tau_l \right) } \right]\, ,
\end{align}
with $\tau_l = a/v_l$. It follows from this that the term involving the bosonic Green's function in Eq.~\eqref{eq:interIT} can be further rewritten as
\begin{align}
 \prod_l e^{2 {g_l^*}^2 \mathcal{G}_l (\tau)} 
 =
  \prod_l e^{2 {g_l^*}^2 \log \left[ \frac{ \sinh \left( i \frac{\pi}{\beta} \tau_l \right) }{\sinh \left( \frac{\pi}{\beta} \left(i \tau_l - \tau\right) \right) } \right]} 
 =
  \prod_l e^{2 {g_l^*}^2 \log \left[ \frac{ \sinh \left( i \frac{\pi}{\beta} \tau_0 \right) }{\sinh \left( \frac{\pi}{\beta} \left(i \tau_0 - \tau\right) \right) } \right] + 2 {g_l^*}^2 \log \left( \frac{\tau_l }{\tau_0 } \right)} 
  =   e^{2 \sum_l {g_l^*}^2 \mathcal{G} (\tau)}   \prod_l \left( \frac{\tau_l }{\tau_0 } \right)^{2 {g_l^*}^2} 
\end{align}
where we used that the short time cutoff $\tau_l$ in the denominator only serves as a regularization and can be replaced by any infinitesimal. This allows to drop the $l$ dependence in the bosonic Green's function and to perform the product over $l$, letting the scaling dimension appear naturally.

The tunneling current can now be rewritten as
\begin{align}
\left\langle I_T (t) \right\rangle = \frac{1}{2} Q_{\mathbf{g}^*}  %\prod_l \left( \frac{\tau_l }{\tau_0 } \right)^{2 {g_l^*}^2} 
 \sum_\epsilon \epsilon \int dt' 
e^{i \epsilon Q_{\mathbf{g}^*} V_\text{DC} (t-t')}
\left[  \Gamma_{\mathbf{g}^*} (t) \right]^\epsilon \left[\Gamma_{\mathbf{g}^*} (t') \right]^{-\epsilon} 
\sum_{\eta \eta'} \eta'  e^{2 \delta_{\mathbf{g}^*} \mathcal{G}^{\eta \eta'} (t-t')} ,
\label{eq:interITfinal}
\end{align}
where, for convenience and without loss of generality, we reabsorbed the prefactor in $\tau_l/\tau_0$ into the definition of the tunneling amplitude. This expression perfectly mirrors the one obtained for the Laughlin case in Eq.~\eqref{eq:avgIB}, where the effective charge $e^*$ and scaling dimension $\nu_D$ of the Laughlin quasiparticle is replaced with the corresponding effective charge $Q_{\mathbf{g}^*}$ and scaling dimension $\delta_{\mathbf{g}^*}$ of the leading tunneling quasiparticle.

%%%%%%%%%%%%%%%%%%%%%%%%%%%%%%%%%%%%%%%%%%%%%%%%%%%%%%%
%    appendix computing current

\section{Computation steps for the current}\label{app:compsteps}
In this section we derive a general formula for the average current Eq.~\eqref{eq:avgIB} without any assumptions on the value of $\nu_D$ other than it being positive. In particular, we obtain Eq.~\eqref{eq:currinter}.

We start by performing the sum over $\eta$ in Eq.~\eqref{eq:avgIB}, using 
\begin{equation}\label{eq:etasum}
	\sum_{\eta,\eta'=\pm} \eta' e^{2 \nu_D \mathcal{G}^{\eta \eta'} \left( t-t' \right)} = 2 \left[ e^{2 \nu_D \mathcal{G} \left( \tau \right)} -e^{2 \nu_D \mathcal{G} \left( -\tau \right)} \right]\Theta(\tau)\, ,
\end{equation}
where $\tau=t-t'$. Then the sum over $\epsilon$ can be performed as
\begin{equation}\label{eq:epssum}
	\sum_{\epsilon=\pm} \epsilon e^{i \epsilon \omega_0^* (t-t')} \left[ \lambda(t) \right]^{\epsilon} \left[ \lambda(t') \right]^{-\epsilon}=\lambda_0^2\sum_{lm}\overline{p_l}p_me^{i(l-m)\Omega t}\left(e^{i(m+q)\Omega\tau} - e^{-i(l+q)\Omega\tau}\right)\, ,
\end{equation}
where $q=\frac{\omega_0^*}{\Omega}$.
Inserting Eq.~\eqref{eq:etasum} and \eqref{eq:epssum} in Eq.~\eqref{eq:avgIB} gives 
\begin{equation}
	\langle I_\text{T} (t) \rangle =  
	e^* \left( \frac{1}{2 \pi a} \right)^2 \lambda_0^2\sum_{l,m}\overline{p_l}p_me^{i(l-m)\Omega t} \int_0^{+\infty} \mathrm{d}\tau\, \left(e^{i(m+q)\Omega\tau} - e^{-i(l+q)\Omega\tau}\right) \left[ e^{2 \nu_D \mathcal{G} \left( \tau \right)} -e^{2 \nu_D \mathcal{G} \left( -\tau \right)} \right]\, .
\end{equation}
The next step is to simplify the expression of the Green's function $\mathcal{G}(\tau)$, see Eq.~\eqref{eq:kgf}, denoting $\eta=\pm$, 
\begin{equation}
		e^{2\nu_D \mathcal{G}(\eta\tau)}=(-i\eta)^{2\nu_D}\tanh\left(\frac{\pi}{\beta}\tau_0\right)^{2\nu_D}\frac{\cosh\left(\frac{\pi}{\beta}\tau\right)^{-2\nu_D}}{\left[\tanh\left(\frac{\pi}{\beta}\tau\right)-i\eta\tan\left(\frac{\pi}{\beta}\tau_0\right)\right]^{2\nu_D}}\, .
\end{equation}
Thus, the current reads
\begin{equation}
	\begin{aligned}
		\langle I_\text{T} (t) \rangle =  
		e^* \left(2v \tau_0\right)^{-2}&\pi^{-3}\beta\lambda_0^2\sum_{l,m}\overline{p_l}p_me^{i(l-m)\Omega t}\sum_{\eta=\pm}\eta(-i\eta)^{2\nu_D}\tanh\left(\frac{\pi}{\beta}\tau_0\right)^{2\nu_D}\\
	&\times\int_0^{+\infty} \mathrm{d}x\, \left[\exp\left(i\frac{m+q}{\pi\theta}x\right) - \exp\left(-i\frac{l+q}{\pi\theta}x\right)\right]\frac{\cosh\left(x\right)^{-2\nu_D}}{\left[\tanh\left(x\right)-i\eta\tan\left(\frac{\pi}{\beta}\tau_0\right)\right]^{2\nu_D}}\, , 	
	\end{aligned}
\end{equation}
where $\theta=\left(\beta\Omega\right)^{-1}$ is the reduced temperature.
Changing variables to $y=\tanh(x)$, one is left with
\begin{equation}
	\begin{aligned}
		\langle I_\text{T} (t) \rangle &=  
		e^* \left(2v\tau_0\right)^{-2}\pi^{-3}\beta\lambda_0^2\sum_{l,m}\overline{p_l}p_me^{i(l-m)\Omega t}\sum_{\eta=\pm}\eta\times\\
		&\times\int_0^{1} \mathrm{d}y\, \left[(1-y)^{\nu_D-1-i\frac{m+q}{2\pi\theta} }(1+y)^{\nu_D-1+i\frac{m+q}{2\pi\theta}}\left(1+i\eta \tan\left(\frac{\pi}{\beta}\tau_0\right)^{-1}y\right)^{-2\nu_D}-(m, q)\to(-l, -q)\right]\, .
	\end{aligned}
\end{equation}
Where the notation $f(a,b) - f(c, d) = f(a,b)- (a, b)\to(c, d)$ has been used.
This integral can be expressed in terms of the first Appell hypergeometric series $F_1$, see Eq.~(3.312) of Ref.~\cite{zwillinger07}, as long as $\nu_D$ is positive.
The current therefore reads 
\begin{equation}
	\begin{aligned}
		\langle I_\text{T} (t) \rangle =  
		e^* &\left(2v\tau_0\right)^{-2}\pi^{-3}\beta\lambda_0^2\sum_{l,m}\overline{p_l}p_me^{i(l-m)\Omega t}\sum_{\eta=\pm}\eta\times\\
		&\Bigg[\text{B}\left(\nu_D-i\frac{m+q}{2\pi\theta}, 1\right)F_1\left(1, 1-\nu_D -i\frac{m+q}{2\pi\theta} , 2\nu_D, 1+\nu_D-i\frac{m+q}{2\pi\theta}; -1; -i\eta \tan\left(\frac{\pi}{\beta}\tau_0\right)^{-1}\right)\\
		&\qquad\qquad\qquad\qquad\qquad\qquad\qquad\qquad\qquad\qquad\qquad\qquad\qquad\qquad\qquad\qquad-(m, q)\to(-l, -q)\Bigg]\, ,
	\end{aligned}
\end{equation}
where $\text{B}(x,y)$ is Euler's Beta function. Using the properties of Appell's hypergeometric series $F_1$, see Eq.~(9.182.1) of Ref.~\cite{zwillinger07}, and the properties of Euler's Beta function, see Eq.~(8.384.1) of \cite{zwillinger07}, we find 
\begin{equation}
	\begin{aligned}
		\langle I_\text{T} (t) \rangle =  
		e^* &\left(2v\tau_0\right)^{-2}\pi^{-3}\beta\lambda_0^2\sum_{l,m}\overline{p_l}p_me^{i(l-m)\Omega t}\\
		\times\sum_{\eta=\pm}\eta&\Bigg[\frac{-i\eta\sin\left(\frac{\pi}{\beta}\tau_0\right)\exp\left(i\eta\frac{\pi}{\beta}\tau_0\right)}{\nu_D-i\frac{m+q}{2\pi\theta}}{_2F_1}\left(1, 1-\nu_D -i\frac{m+q}{2\pi\theta} ; 1+\nu_D-i\frac{m+q}{2\pi\theta}; \exp\left(2i\eta\frac{\pi}{\beta}\tau_0\right)\right)\\
		&\qquad\qquad\qquad\qquad\qquad\qquad\qquad\qquad\qquad\qquad\qquad\qquad\qquad\qquad\qquad\qquad-(m, q)\to(-l, -q)\Bigg]\, ,
	\end{aligned}
\label{eq:generalIT}
\end{equation}
where $_2F_1$ is Gauss' hypergeometric function.
We stress that this formula is valid to all orders in $\tau_0$ and for any $\nu_D>0$.

%%%%%%%%%%%%%%%%%%%%%%%%%%%%%%%%%%%%%%%%%%%%%%%%%%%%%%%%%%
%     appendix weak backscattering

\section{Weak backscattering regime}\label{app:wbcurrent}
In this appendix, we find a formula for the current as a function of time in the regime where $0<\nu_D<1$, with $\nu_D \neq 1/2$. The weak backscattering regime for Laughlin fractions is a particular case of this regime of parameters. Therefore, performing an expansion in $\tau_0/\beta$ we derive Eq.~\eqref{eq:avIwb}.

In this regime, as neither $-\nu_D$, nor $2\nu_D-1$ are integer, and as $\left\lvert 1-\exp\left(2i\eta\frac{\pi}{\beta}\tau_0\right)\right\rvert<\pi$, Eq.~(2.10.1) of Ref.~\cite{bateman53} holds.
Thus, to lowest order in $\tau_0/\beta$
\begin{equation}\label{eq:batemanformula}
	\langle I_\text{T} (t) \rangle = \frac{\Gamma(2\nu_D)\mathcal{I}}{2} \sum_{l,m}\overline{p_l}p_me^{i(l-m)\Omega t}\sum_{\eta=\pm}(i\eta)^{2\nu_D}\left[\frac{\Gamma(1-2\nu_D)\Gamma\left(1+\nu_D-i\frac{l+q}{2\pi\theta}\right)}{\left(\nu_D-i\frac{m+q}{2\pi\theta}\right)\Gamma\left(1-\nu_D-i\frac{m+q}{2\pi\theta}\right)}-(m, q)\to(-l, -q)\right]\, ,
\end{equation}
where $\Gamma(z)$ is the Gamma function, $\Lambda=(\Omega\tau_0)^{-1}$ and 
\begin{equation}
    \mathcal{I}=\frac{e^*\Omega}{\pi}\left(\frac{\lambda_0}{v}\right)^2\left(\frac{2\pi\theta}{\Lambda}\right)^{2\nu_D-2}\frac{\theta}{\Gamma(2\nu_D)}\, .
\end{equation}
Note that for $1/2 < \nu_D <1$, the expansion of the hypergeometric function leads to a leading contribution of order $O (\tau_0)$ which dominates over the one considered above, which is of order $O (\tau_0^{2\nu_D})$. However, this leading contribution simplifies when accounting for the second hypergeometric function obtained upon exchanging $(m, q)\to(-l, -q)$.

Using the fact that $i\eta=\exp\left(i\eta\frac{\pi}{2}\right)$, Euler's reflection formula for the Gamma function [see \cite{zwillinger07}, Eq.~(8.384.1)] and the fact that $\Gamma(z^*)=\Gamma(z)^*$ [which can be deduced from \cite{zwillinger07}, Eq.~(8.334.3)], the current is reduced to
\begin{equation}
	\langle I_\text{T} (t) \rangle = -i\frac{\mathcal{I}}{2\cos(\pi\nu_D)} \sum_{l,m}\overline{p_l}p_me^{i(l-m)\Omega t}\left[\left\lvert\Gamma\left(\nu_D-i\frac{m+q}{2\pi\theta}\right)\right\rvert^2\sin\left(\pi\nu_D+i\frac{m+q}{2\theta}\right)-(m, q)\to(-l, -q)\right]\, .
\end{equation}
Finally, after a change of indices, one is left with
\begin{equation}\label{eq:avgIBgen}
	\langle I_\text{T} (t) \rangle =  
		\mathcal{I}\sum_{l,m}\left\lvert\Gamma\left(\nu_D+i\frac{m+q}{2\pi\theta}\right)\right\rvert^2
		\Im\left\{\overline{p_m}p_{m-l}e^{il\Omega t}\left[\tan\left(\pi\nu_D\right)\cosh\left(\frac{m+q}{2\theta}\right) +i \sinh\left(\frac{m+q}{2\theta}\right)\right]\right\}\, ,
\end{equation}
where $\Im(x)$ denotes the imaginary part of $x$.
Assuming that $p_l\in\mathds{R}$, this reduces to 
\begin{equation}
	\begin{aligned}
		\left\langle I_\text{T} (t) \right\rangle =\mathcal{I} \sum_{l>0}\Bigg[&\cos(l\Omega t)\sum_{m}\left\lvert\Gamma\left(\nu_D+i\frac{m+q}{2\theta\pi}\right)\right\rvert^2(p_{m-l}p_m + p_m p_{l+m} )\sinh\left(\frac{m+q}{2\theta}\right)\\
		+&\sin(l\Omega t)\tan(\pi\nu_D)\sum_{m}\left\lvert\Gamma\left(\nu_D+i\frac{m+q}{2\theta\pi}\right)\right\rvert^2( p_mp_{l+m} -p_{m-l}p_m )\cosh\left(\frac{m+q}{2\theta}\right)\Bigg]\\
		+& \mathcal{I} \sum_{m}\left\lvert\Gamma\left(\nu_D+i\frac{m+q}{2\theta\pi}\right)\right\rvert^2p_m^2\sinh\left(\frac{m+q}{2\theta}\right)\, .
\end{aligned}
\end{equation}
The current can finally be rewritten as
\begin{equation}\label{eqapp:avIwb}
	\left\langle I_\text{T}(t) \right\rangle = I_0 + \sum_{n>0} \mathcal{I} C_n\cos(n\Omega t + \varphi_n)\, ,
\end{equation}
where
\begin{equation}\label{eq:fcoefwb}
	\begin{aligned}
		\varphi_n &=\arctan\left(\frac{B_n}{A_n}\right)\\
		A_l &= \sum_m\left\lvert\Gamma\left(\nu_D+i\frac{m+q}{2\theta\pi}\right)\right\rvert^2(p_{m-l}p_m + p_{l+m}p_m)\sinh\left(\frac{m+q}{2\theta}\right)\\
		B_l &= \tan(\pi\nu_D)\sum_m\left\lvert\Gamma\left(\nu_D+i\frac{m+q}{2\theta\pi}\right)\right\rvert^2(p_{m-l}p_m - p_{l+m}p_m ) \cosh\left(\frac{m+q}{2\theta}\right)\\
		C_n&=\frac{A_n}{\cos(\varphi_n)}\\
		I_{0} &= \mathcal{I} \sum_{m}\left\lvert\Gamma\left(\nu_D+i\frac{m+q}{2\theta\pi}\right)\right\rvert^2p_m^2\sinh\left(\frac{m+q}{2\theta}\right)\, .
	\end{aligned}
\end{equation}

%%%%%%%%%%%%%%%%%%%%%%%%%%%%%%%%%%%%%%%%%%%%%%%%%%%%%%%%%%%%
%   appendix Fermi liquid

\section{Fermi liquid calculation} \label{app:fermiliq}

In this appendix, we provide some details of the derivation of the current in the Fermi liquid picture. Or starting point is the expression for the average current in terms of the Keldysh Green's function for electron operators, namely
\begin{equation}
    \left\langle I_\text{T}(t)\right\rangle = -e\left[\lambda(t)G_{\text{RL}}^{+-}(t,t) - \lambda^*(t)G_{\text{LR}}^{+-}(t,t)\right]\, .
\end{equation}
To leading order in the tunnel coupling $\lambda$, using Dyson equation, the dressed Green's function reads
\begin{equation}
    G_{ss'}^{+-}(t,t)=\int\mathrm{d}t \left[ g_{\text{ss}}^{+-}(t-t')\lambda^*(t')g_{\text{s's'}}^{a}(t-t')  + g_{\text{ss}}^{r}(t-t')\lambda^*(t')g_{\text{s's'}}^{+-}(t-t') \right] \, .
\end{equation}
where $g_{ss'}^{\eta\eta'} (t)$ are the bare Keldysh Green's functions (in the absence of tunneling), and $g_{ss'}^{r/a} (t) = g_{ss'}^{++} (t) - g_{ss'}^{\pm \mp} (t)$.

Going to Fourier space, i.e., performing double Fourier transform and keeping in mind that we compute the Green's function at simultaneous time, we can write
\begin{align}
   G_{\text{RL}}^{+-}(t,t) &= \lambda_0 \sum_l \overline{p_l}e^{il \Omega t}\int\frac{\mathrm{d}\omega}{2\pi} \left[ g_{\text{RR}}^{+-}(\omega) g_{\text{LL}}^{a} (\omega+l\Omega) + g_{\text{RR}}^{r} (\omega) g_{\text{LL}}^{+-} (\omega+l\Omega) \right] \\
   G_{\text{LR}}^{+-}(t,t) &= \lambda_0 \sum_l p_l e^{-il \Omega t} \int \frac{\mathrm{d}\omega}{2\pi} \left[ g_{\text{LL}}^{+-}(\omega) g_{\text{RR}}^{a} (\omega+l\Omega) + g_{\text{LL}}^{r} (\omega) g_{\text{RR}}^{+-} (\omega+l\Omega) \right]\, .
\end{align}
The current can then be readily written as
\begin{equation}
\begin{aligned}
    \left\langle I_\text{T}(t)\right\rangle = -e\lambda_0^2\sum_{l,m} \overline{p_l}p_me^{i(l-m) \Omega t}\int\frac{\mathrm{d}\omega}{2\pi}\Big[&g_{\text{RR}}^{+-}(\omega)g_{\text{LL}}^{a}(\omega+l\Omega) + g_{\text{RR}}^{r}(\omega)g_{\text{LL}}^{+-}(\omega+l\Omega)\\
    &-g_{\text{LL}}^{+-}(\omega)g_{\text{RR}}^{a}(\omega+m\Omega) - g_{\text{LL}}^{r}(\omega)g_{\text{RR}}^{+-}(\omega+m\Omega)\Big]\, .
\end{aligned}
\end{equation}
In order to (later on) make a correspondence with the $\nu_D=1$ limit of the chiral Luttinger liquid calculation, we need to make consistent assumptions between the two models. In the present case, this means that we have to work in the wide band limit.
This limit is implemented by setting
\begin{equation}
\begin{aligned}
    g_{\text{ss}}^{r/a} (\omega) &=\mp i(2v_\text{F})^{-1}\\
    g_{\text{ss}}^{+-}(\omega)&=2i\Im\left[g_{\text{ss}}^{a} (\omega) \right] f(\omega-\mu_{s})\, ,
\end{aligned}
\end{equation}
with $\mu_\text{s}$ the chemical potential on side $s$ and $f(x)$ the Fermi distribution.
The current can therefore be simplified into
\begin{equation}
    \left\langle I_\text{T}(t)\right\rangle = \frac{\lambda_0^2 e}{4\pi v_{\text{F}}^2}\sum_{l,m} \overline{p_l}p_me^{i(l-m) \Omega t}\int\mathrm{d}\omega\Big[f(\omega-\mu_\text{R}) - f(\omega+l\Omega-\mu_\text{L})-f(\omega-\mu_\text{L}) + f(\omega+m\Omega-\mu_\text{R})\Big]\, .
\end{equation}
Performing the integration yields
\begin{equation}
    \left\langle I_\text{T}(t)\right\rangle = \frac{e\Omega\lambda_0^2}{4\pi v_F^2}\sum_{l,m} \overline{p_l}p_me^{i(l-m) \Omega t}(2q+l+m)\, .
\end{equation}

%%%%%%%%%%%%%%%%%%%%%%%%%%%%%%%%%%%%%%%%%%%%%%%%%%%%%%%%%%%%
%   appendix strong backscattering

\section{Strong backscattering regime}\label{app:sbcurrent}
In this appendix, we derive a formula for the current in the strong backscattering regime. Applying the duality transformation to Eq.~\eqref{eq:generalIT} taken for the Laughlin series, one has for the current in this regime
\begin{equation}
	\begin{aligned}
		\langle I_\text{T} (t) \rangle =  
		e &\left(2v\tau_0\right)^{-2}\pi^{-3}\beta\lambda_0^2\sum_{l,m}\overline{p_l}p_me^{i(l-m)\Omega t}\\
		\times\sum_{\eta=\pm}\eta&\Bigg[\frac{-i\eta\sin\left(\frac{\pi}{\beta}\tau_0\right)\exp\left(i\eta\frac{\pi}{\beta}\tau_0\right)}{\nu^{-1}-i\frac{m+q}{2\pi\theta}}{_2F_1}\left(1, 1-\nu^{-1} -i\frac{m+q}{2\pi\theta} ; 1+\nu^{-1}-i\frac{m+q}{2\pi\theta}; \exp\left(2i\eta\frac{\pi}{\beta}\tau_0\right)\right)\\
		&\qquad\qquad\qquad\qquad\qquad\qquad\qquad\qquad\qquad\qquad\qquad\qquad\qquad\qquad\qquad\qquad-(m, q)\to(-l, -q)\Bigg]\, ,
	\end{aligned}
\end{equation}
where $q = \omega_0/\Omega$.

We now want to perform an expansion in low $\tau_0/\beta$ for the Laughlin series (where $\nu^{-1}$ is an odd integer). The technique used in the case $0<\nu_D<1$ (see Appendix~\ref{app:wbcurrent}) cannot be used here as $1-2\nu^{-1}$ is an integer, thus, Eq.~(2.10.1) of Ref.~\cite{bateman53} does not hold. 
However, another route is possible and, as we will show, the leading contribution is of first order or less in $\tau_0$. 

Following Prudnikov et. al. [see Eq.~(7.3.1.31) of Ref.~\cite{prudnikov90}] we perform a logarithmic expansion of the hypergeometric function and remove terms of order higher than three. The current can then be written as the sum of two terms:
\begin{equation}\label{eq:logexpansion}
    \langle I_\text{T}(t) \rangle = \langle I_\text{Fermi/DC}(t) \rangle + \langle I_\text{cor-AC}(t) \rangle\, ,
\end{equation}
where the term containing the DC behavior as well as the Fermi limit is
\begin{equation}\label{eq:dcfermi}
	\begin{aligned}
		\langle I_\text{Fermi/DC} (t) \rangle \approx  
		ie &\left(2v\tau_0\right)^{-2}\pi^{-3}\beta\lambda_0^2\sum_{l,m}\overline{p_l}p_me^{i(l-m)\Omega t}\sum_{\eta=\pm}\sin\left(\frac{\pi}{\beta}\tau_0\right)\frac{\Gamma\left(\nu^{-1}-i\frac{m+q}{2\pi\theta}\right)}{\Gamma\left(1-\nu^{-1}-i\frac{m+q}{2\pi\theta}\right)}\\
		\times\sum_{k,r=0}^\infty\sum_{s=0}^{2\nu^{-1}-1+k}\Bigg\{&(-1)^k2^{k+2\nu^{-1}-1}(i\eta)^{2k-s-2}\frac{(k+2\nu^{-1}-1)_k(2\nu^{-1})_k\left(\nu^{-1}-i\frac{m+q}{2\pi\theta}\right)_k\left(\frac{\pi}{\beta}\tau_0\right)^{2k-s+4\nu^{-1}-2}}{r!s!(k+2\nu^{-1}-1-s)!}\\
		&\qquad\times \left[\log\left(2\frac{\pi}{\beta}\tau_0\right) + i\eta \frac{\pi}{2}-\psi(k+1)+\psi\left(\nu^{-1}+k-i\frac{m+q}{2\pi\theta}\right)\right]-(m, q)\to(-l, -q)\Bigg\}
		\, ,
			\end{aligned}
\end{equation}
while the term corresponding to the AC current when correlations are present reads
\begin{equation}\label{eq:accor}
\begin{aligned}
		\langle I_\text{cor-AC}(t) \rangle \approx  
		-ie &\left(2v\tau_0\right)^{-2}\pi^{-3}\beta\lambda_0^2\sum_{l,m}\overline{p_l}p_me^{i(l-m)\Omega t}\sum_{\eta=\pm}\sin\left(\frac{\pi}{\beta}\tau_0\right)\frac{1}{2\nu^{-1}-1}\\
		\times&\sum_{r=0}^\infty\sum_{k=0}^{2\nu^{-1}-2}\sum_{s=0}^k\Bigg[(-1)^k(i\eta)^{r+2k-s}2^{k}\frac{(1)_k (1-\nu^{-1}-i\frac{m+q}{2\pi\theta})_k\left(\frac{\pi}{\beta}\tau_0\right)^{r+2k-s}}{r!s!(k-s)!(2-2\nu^{-1})_k}-(m, q)\to(-l, -q)\Bigg]\, ,
\end{aligned}
\end{equation}
with $(x)_n=\prod_{k=0}^{n-1}(x+k)$ is the Pochhammer symbol.

One sees indeed that for $\nu=1$ Eq.~\eqref{eq:accor} vanishes and for $\nu^{-1}>1$ all terms in Eq.~\eqref{eq:dcfermi} are sub-leading.
Therefore, we present the computation in two different sections.

%%%%%%%%%%%%%%%%%%%%%%%%%%%%%%%%%%%%%%%%%%%%%%%%%%%%%%

\subsection{Fermi liquid limit, \texorpdfstring{$\nu=1$}{nu=1}}

The Fermi liquid limit is obtained by setting $\nu^{-1}=\nu=1$.
The first sum over $k$ in Eq.~\eqref{eq:accor} vanishes and only the terms of Eq.~\eqref{eq:dcfermi} contribute.
Performing the sum over $\eta$ removes all terms containing odd powers of $\eta$, i.e., to lowest order in $\tau_0/\beta$ ($s=1$, $k=r=0$) the current reads
\begin{equation}\label{eq:luttingerfermiliquid}
		\langle I_\text{T} (t) \rangle = \frac{e\lambda_0^2\Omega}{4\pi v_{\text{F}}^2}\sum_{l,m}\overline{p_l}p_me^{i(l-m)\Omega t}(2q+l+m)\, .
\end{equation}
%Finally, injecting Eq.~(\ref{eq:lambdadef}) we find
%\begin{equation}
%    	\frac{e^2 }{2\pi v_{\text{F}}^2}\left\{{\begin{aligned}
%			\lambda(t)^2V_{\text{dc}}&\qquad\text{for a gate drive}\\
%			\lambda_0^2V(t)&\qquad\text{for a voltage drive}
%		\end{aligned}}\right.
%\end{equation}
%which corresponds to the standard Fermi liquid result, see App.~\ref{app:fermiliquid}.

%%%%%%%%%%%%%%%%%%%%%%%%%%%%%%%%%%%%%%%%%%%%%%%%%%%%%%%%%%%

\subsection{Filling factors, \texorpdfstring{$\nu^{-1}>1$}{v*>1}}
The general case of integer $\nu^{-1}$ greater than one is obtained by remarking that the leading term in Eq.~\eqref{eq:dcfermi} is a polynomial of order five or more in $\tau_0/\beta$. 
Therefore, we can extract the main contribution to the current from Eq.~\eqref{eq:accor} only.
It can be shown that the term of order minus one in $\tau_0/\beta$ does not depend on $q$ nor $l$ or $m$, thus it does not contribute in the current.
The sum over $\eta$ removes the zeroth order term in $\tau_0/\beta$ so we are left with the first order contributions.
They arise when the indices respect $r+2k-s=2$, i.e., there are four possibilities:
\begin{itemize}
	\item $r=0$, $k=s=2$;
	\item $r=0$, $k=1$, $s=0$;
	\item $r=k=s=1$;
	\item $r=2$, $k=s=0$.
\end{itemize}
We remark that the last possibility gives rise to a term independent of $q$, $l$, $m$ which thus vanishes. 
Selecting the three first possibilities only and removing the Fermi/DC term in Eq.~\eqref{eq:logexpansion}, the current ultimately reads 
\begin{equation}\label{appeq:avIsb}
		\langle I_\text{T} (t) \rangle\approx\frac{-e}{\pi^2 (1-2\nu^{-1})_3}\left(\frac{\lambda_0}{v}\right)^2\frac{\Omega}{\Lambda}\Im\left[\sum_{l,m}\overline{p_l}p_me^{i(l-m)\Omega t}\left(m+q\right)^2\right]\, .
\end{equation}

\end{appendix}

\twocolumngrid
\bibliography{fractional-quantum-hall-effect-gate-modulation}
\end{document}